\documentclass{amsart}
\usepackage{style}

\title[
    Enhancing VQAs for
    Multicriteria Optimization
]{
    Enhancing Variational Quantum Algorithms for
    Multicriteria Optimization
}

\author{Ivica Turkalj}
\email{ivica.turkalj@itwm.fraunhofer.de}
\author{Tom Ewen}
\email{tom.ewen@itwm.fraunhofer.de}
\author{Pascal Halffmann}
\email{pascal.halffmann@itwm.fraunhofer.de}
\author{Janik Maciejewski}
\email{janik.maciejewski@ruv.de}
\author{Michael Trebing}
\email{michael.trebing@itwm.fraunhofer.de}
\author{Zakaria Abdelmoiz Dahi}
\email{abdelmoiz-zakaria.dahi@inria.fr}

\address[I. Turkalj,T. Ewen, P. Halffmann, M. Trebing]{Fraunhofer ITWM}
\address[J. Maciejewski]{R+V Lebensversicherung AG}
\address[Z. Dahi]{Inria centre, University of Lille}

\thanks{This research was funded by Bundesministerium für Bildung und Forschung (BMBF),
grant number 13N15695.}

\date{\today}

\begin{document}

\begin{abstract}
    This paper presents methodological improvements to variational quantum algorithms (VQAs) for 
    solving multicriteria optimization problems. 
    We introduce two key contributions. 
    First, we reformulate the parameter optimization task of VQAs as a multicriteria problem, 
    enabling the direct use of classical algorithms from various multicriteria 
    metaheuristics. 
    This hybrid framework outperforms the corresponding single-criteria VQAs 
    in both average and worst-case performance across diverse benchmark problems. 
    Second, we propose a method that augments the 
    hypervolume-based cost function with coverage-oriented indicators, 
    allowing explicit control over the diversity of the resulting Pareto front approximations. 
    Experimental results show that our method can improve coverage by up to 40\% 
    with minimal loss in hypervolume. 
    Our findings highlight the potential of combining quantum variational methods 
    with classical population-based search to advance practical quantum optimization.
\end{abstract}

\maketitle

\section{Introduction}
Quantum computing, by leveraging quantum mechanical phenomena such as superposition and entanglement, has demonstrated significant promise in solving complex computational problems. In particular, complex optimization problems, such as those labeled as NP-hard have been at the forefront at potential application areas, where quantum computing can solve these problems more efficient than classical resolution strategies~\cite{AuYeung2023QuantumOptimization}. Among various quantum computing paradigms such as quantum annealing, variational quantum algorithms (VQAs) have emerged as a particularly compelling approach due to their hybrid quantum-classical nature, allowing current quantum devices, often limited by noise and coherence times, to solve meaningful problems. These algorithms iteratively adjust the parameters of quantum circuits (ansatz) to minimize or maximize objective functions, which are evaluated through quantum computations and optimized classically. Several studies observe advantages such as a better scaling, that is a better performance for large instance sizes~\cite{Shaydulin2024ScalingAdvantage}, or better approximation quality~\cite{Pirnay2024ApproximationAdvantage}.
However, a true performance advantage in practice, either in solution time or quality has not been shown yet. Indeed, some publications hint that the limitations of current quantum optimization algorithms as QAOA prohibited any quantum advantage~\cite{Mueller2025Limits} and claims of quantum supremacy of quantum annealing have quickly been debunked~\cite{Mauron2025Anti-DWave}.

A particular reason is the vast performance of current optimization solvers such as Gurobi that concentrate over 60 years of expertise in solving optimization problems in one solver framework. Therefore, benefits from applying quantum algorithms may arise first for complex optimization problems, for which high-performance classical solvers are lacking. Multicriteria optimization problems that deal with optimizing several, often conflicting objective function simultaneously are one of the most complex optimization problems. Besides their NP-hardness, there does not one optimal objective function value but several compromise solutions, so-called efficient or Pareto-optimal solutions with nondominated objective functions value exists. Finding all these compromises is computational demanding and even intractable due to the exponential (w.r.t. the encoding length of the problem) number of these compromises. While there exist many classical algorithms to tackle these problems, both exact methods~\cite{Halffmann2022Survey} and heuristics such as the famous Non-dominated Sorting Genetic Algorithm II (NSGAII)~\cite{nsga2} that is incorporated in the pymoo Python package~\cite{Blank2020Pymoo}, these algorithms are by far not as performant as solvers for e.g., mixed-integer linear problems. Since multicriteria optimization problems are omnipresent in practice, there is high need for faster and better multicriteria optimization algorithms and a chance for quantum algorithms to make a substantial impact, as it has been stated in the seminal overview on quantum optimization by Abbas et al.~\cite{Abbas2024Overview}.

\subsection{Recent Research}
There already exist several publications in the overlap of quantum computing and multicriteria optimization. Most of these publications are focused on the development of quantum-inspired algorithms to tackle multicriteria optimization problems, in particular heuristics such as evolutionary algorithms that take inspiration from quantum effects~\cite{Kim2006Quantuminspired}. As these algorithms are solely classical approaches, we omit these in this overview.

Multicriteria optimization has also been identified as a tool to improve quantum algorithms and their execution on real hardware. In particular, compiling a quantum circuit to the actual hardware while respecting the hardware architecture requirements is a multicriteria problem itself, e.g. one wants to minimize the circuit depth while minimize the number of ancilla qubits or error rates. Several papers have identified this multicriteria problem and provided multicriteria heuristics to improve the compiling, e.g. \cite{Baran2021Compiling,Ruffinelli2017Compiling, Swierkowska2024Compiling}. Further, the paper \cite{Diez2023} used a multicriteria view on the optimization of the parameters of variational algorithms for the resolution of optimization problems. They utilize NSGA-II to optimize the variational parameters subject to two objectives: the first one is the original objective of the optimization problem. The second objective is a function that evaluates the feasibility of the solution, i.e., it comprises of the penalty functions deduced from the constraints of the optimization problem.

In a recent years,  quantum methods also have been developed to directly target multicriteria problems. The first method has been published in 2016~\cite{Baran2016Annealing} and transforms the multicriteria problem into a single-criterion problem via weighted sum and solve this problem using D-Wave's quantum annealer. Similar approaches using either QAOA or annealing have been presented for specific problems in \cite{Chiew2024NetworkRouting} for network routing and for portfolio optimization~\cite{Aguilera2024PortfolioOptimization, Turkalj2024PortfolioOptimization}.
Similarly, in \cite{Urgelles20226GCommunication} the multicriteria problem is transformed into a single-criterion problem, the so-called lexicographic problem, where one objective is optimized and then added as constraint when optimizing the second objective function. The resulting single-criterion problems are solved using QAOA. In \cite{Baran2018Grover}, the authors present a variant of Grover adaptive search, where the oracle flags solutions whose objective function values are not dominated. Given a starting point, this approach then generates an efficient solution.

So far, all these approaches only provide a single efficient solution to the multicriteria problem. 
The first algorithms that aims at finding the whole Pareto front is presented in \cite{Eks2024}. 
The authors propose a variant of QAOA for qudits instead of qubits, where for each objective function a 
problem Hamiltonian and a mixer Hamiltonian are incorporated in one layer of the QAOA. 
The variational parameters are then optimized with respect to the hypervolume of the obtained shots 
from the QAOA circuit execution. 
We will investigate this apporach in more detail in Section~\ref{sec:variational}.

The authors in \cite{zak} propose a decomposition-based Multi-Objective QAOA (MO-QAOA)
using weighted sum 
and Tchebycheff scalarisation, while accounting for the constraints of current 
quantum hardware and simulators.

Most recently, \cite{kotil2025} revisit the approach of \cite{Baran2016Annealing}. They use the resolution of the weighted sum problem as sampling device for a heuristic to find the whole Pareto front. Basically, they provide a QAOA circuit with pre-trained variational parameters that work for arbitrary weighted sum weights. By executing this circuit for randomly sample weights and filtering the efficient solutions, they show that this approach is compatible to some classical multicriteria solvers in speed and accuracy.

\subsection{Our Contribution}
By investigating the algorithm presented in~\cite{Eks2024}, we introduce two appealing directions of improving this algorithm. 

It is somewhat counter-intuitive to use the hypervolume as a single metric to optimize the variational parameters, when the underlying optimization problem is a multicriteria one, since for single-criterion problems, we also use the original objective function as evaluation metric. Therefore, we introduce a novel framework integrating classical multicriteria optimization algorithms into the optimization of variational quantum circuits. We show how to interpret the quantum state, generated by the parametrized ansatz, as the objective of a multicriteria optimization problem, which is then accessible for any algorithm for multicriteria optimization to optimize the variational parameters. In particular, we utilize this framework to apply NSGA-II as classical solver to the quantum circuit presented in~\cite{Eks2024}.

If one wants to rely on single-criterion algorithms (e.g., due to potentially better running time), we also introduce more sophisticated evaluation metrics to the evaluation of the measurement of the quantum circuit. In particular, we enhance the existing hypervolume by adding a metric that measure the coverage of the Pareto front such as the Pareto spread. This is crucial for ensuring solution diversity and 
improving coverage along the Pareto front. Our study explicitly compares different coverage indicators, 
evaluating their ability to enhance the quality, diversity, 
and practical relevance of solutions produced by the hybrid quantum-classical algorithm.

Additionally, we present a comprehensive experimental study that evaluates our proposed methodology across several benchmark problems. Our experiments demonstrate that using NSGA-II as classical optimizer consistently achieves improved performance (up to 10\%) and increased solution stability in comparison to the algorithm in~\cite{Eks2024}. Further, for our second approach, the experiments show that we can obtain a significantly (up to 40\%) better coverage by giving up no to little hypervolume.

The remainder of this article is structured as follows: In the next section, we introduce notation and concepts of multicriteria optimization. In Section~\ref{sec:variational}, we provide a detailed description of the algorithm presented in~\cite{Eks2024}. Section~\ref{sec:qnsgaii} advances this algorithm with an integration of multiobjective heuristics as classical solvers for the variational parameters. This is followed with a specific integration of NSGA-II in Section~\ref{sec:qnsgaii}. Next, we provide another extension by incoporating coverage metrics in the evaluation of the current sample set in Section~\ref{sec:spread}. A computational study for these two advancements is given in Section~\ref{sec:experiments}. We conclude this article in Section~\ref{sec:conclusion}.
\section{Multicriteria Optimization}\label{sec:mco}

In this paper, we consider unconstrained, binary, multicriteria optimization problems. 
We refer to \cite{Ehr2005} as a general point of reference.

Let $\F = \menge{0,1}$ be the field with two elements.
Throughout the paper, $n \in \N$ will denote the number of decision variables 
and $K \in \N$ the number of objective functions.
We consider criteria of the form
\begin{align*}
    f_k: \F[n] \longrightarrow \R,
\end{align*}
where $k=1, \ldots, K$.
The vector-valued function, that is composed of the $f_k$, is written as
\begin{align*}
    f: \F[n] \longrightarrow \R^K, \quad x \longmapsto (f_1(x), \ldots, f_K(x)).
\end{align*}
The vector space $\F[n]$ is referred to as the \emph{decision space} 
and the vector space $\R^K$ as the \emph{objective space}.
We typically denote subsets of the decision space with the symbol $\X$ 
and subsets of the objective space with $\Y$.
To define the optimization tasks, we note that the relation 
\begin{align*}
    y^1 \leqq y^2 :\iff  y_k^1 \leq y_k^2 \quad \forall\, k=1,\ldots,K,
\end{align*}
where $y^1, y^2 \in \R^K$, defines a partial order on $\R^K$.
For a subset $\Y \subseteq \R^K$, a \emph{minimal element} of $\Y$ (with respect to $\leqq$) is 
every element $y^1 \in \Y$ with the following property:
\begin{align*}
    y^2 \leqq y^1, y^2 \in \Y \implies y^2=y^1.
\end{align*}
That is, $y^1$ is minimal if no element in $\Y$ is 
smaller than $y^1$ with respect to $\leqq$.
The set of all minimal elements of $\Y$ is denoted as 
\begin{align*}
    \min \Y := \Menge{y \in \Y}{\text{$y$ is a minimal element of $\Y$}}.
\end{align*}
Generally, a set has multiple minimal elements.

Given the partial order on $\R^K$, we can use $f$ to define a corresponding partial order on $\F[n]$,
\begin{align*}
    x^1 \preceq_f x^2 :\iff f(x^1) \leqq f(x^2),
\end{align*}
where $x^1, x^2 \in \F[n]$. Minimal elements of $\F[n]$ are defined analogously.
The relations $\leqq$ and $\preceq_f$ are also called \emph{dominance relations}.
Elements of the set $\min \F[n]$ are called \emph{efficient solutions}, and its 
image $f(\min \F[n])$ is called \emph{Pareto front}.
Elements in the Pareto front are also called \emph{nondominated}.
From the definitions it immediately follows that $f(\min \F[n]) = \min f(\F[n])$.

With this preparation, we define an unconstrained binary multicriteria optimization problem as 
the task of solving any of the following:
\begin{description}
    \item[(MCO1)] Determine $\min \F[n]$ with respect to $\preceq_f$.
    \item[(MCO2)] Determine a set $\X \subseteq \F[n]$ such that $f(\X$) approximates 
    the Pareto front $\min f(\X)$, 
    whereby a suitable measure of proximity between subsets of $\R^K$ 
    has to be specified in advance.
\end{description}
For the latter problem, there exist multicriteria approximation 
algorithms that provide a set of solutions that approximate the Pareto front 
for a given (multi-dimensionnal) approximation factor. 
We refer to \cite{mco_approximation} 
for an overview on definitions and methods. 
As state-of-the-art variational algorithms act as heuristics for the 
resolution of optimization problems, 
we focus on well-established proximity measures for heuristics. 
One of the most frequently used measures in this area is the hypervolume indicator
\cite{Zit1998}.  
Given a set $\Y \subseteq \R^K$ and a reference point $r \in \R^K$, 
the \emph{hypervolume indicator} of $\Y$ with respect to $r$ is defined as
\begin{align*}
    \HV(\Y):= \lambda_K(\Menge{z \in \R^K}{\text{$z \leqq r$ and $\exists\, y \in \Y$ with $y \leqq z$}}),
\end{align*}
where $\lambda_K$ is the Lebesgue measure on $\R^K$.
In other words, the hypervolume indicator measures how much volume is enclosed between $\Y$ and the reference point $r$.
Thus, the set $f(\X)$ is considered close to the set $\min f(\F[n])$, iff the real number $\HV(f(\X))$ is close to 
the real number $\HV(\min f(\F[n]))$.
We will discuss other proximity measures later in \Cref{sec:spread} 
which are designed to emphasize certain properties of sets, 
such as the distribution of its points.
We omit the dependence of the hypervolume on the reference point in the notation,
as it is always chosen canonically based on the 
problem, specifically, the approximate nadir $\tilde{y}^N$ defined below.
Since other proxmity measures also depend on certain auxiliary vectors, 
we collect below all the relevant objects in this regard.

Let $x^{(k)} \in \F[n]$ denote the solution to the single-objective optimization problem associated to $f_k$,
\begin{align*}
    x^{(k)} := \argmin_{x \in \F[n]} f_k(x).
\end{align*}
The \emph{ideal} $y^I \in \R^K$ is the vector in objective space whose entries are the optimal values associated to the 
optimal solutions $x^{(k)}$,
\begin{align*}
    y_k^I := f_k(x^{(k)}).
\end{align*}
The \emph{nadir} $y^N \in \R^K$ is the vector in objective space whose entries are given by 
\begin{align*}
    y^N_{k} := \max_{x \in \min \F[n]} f_k(x).
\end{align*}
Calculating the nadir is generally difficult \cite{Ehr2005}, 
therefore the following approximation $\tilde{y}^N \in \R^K$ to the nadir is 
considered,
\begin{align*}
    \tilde{y}_k^N := \max \menge{f_k(x^{(1)}), \ldots, f_k(x^{(K)})}.
\end{align*}
For multicriteria problems with two objectives, 
it holds that $y^N = \tilde{y}^N$ \cite{Ehr2005}.
Referring back to the hypervolume indicator, 
we will always use $\tilde{y}^N$ as the reference point.

To conclude this section, we briefly discuss the computational 
complexity of binary multicriteria optimization problems.
For a broader description, please refer to \cite{Ehr2000}.

Given a binary multicriteria optimization problem,
it is generally not possible to solve \textbf{MCO1}
in polynomial-time.
For some problem instances, this follows already from the fact that the number of elements in 
$\min \F[n]$ is exponential in the size of the problem instance.
Consider, for example, the following bicriteria problem instance,
\begin{align*}
    f_1(x)  &= \sum_{i=1}^n 2^{i-1}  x_i, \\
    f_2(x)  &= \sum_{i=1}^n (-1) 2^{i-1}  x_i.
\end{align*}
The image 
\begin{align*}
    f(\F[n]) = \menge{
        \begin{pmatrix}
            0 \\ 0
        \end{pmatrix},
        \begin{pmatrix}
            1 \\ -1
        \end{pmatrix},
        \ldots,
        \begin{pmatrix}
            2^n - 1 \\ -2^n -1
        \end{pmatrix}
    }
\end{align*}
lies on the line through the origin in $\R^2$ with slope $-1$, so every point in $f(\F[n])$ is minimal.

Even if the cardinality of $\min \F[n]$ is small, there is generally no efficient algorithm for determining its elements.
Consider the following problem as a generalization of the above,
\begin{align}\label{eq:UMOCO_complexity}
    \tag{UMOCO}
\begin{split}
    f_1(x)  &= \sum_{i=1}^n c_i^1  x_i, \\
            &\vdots \\
    f_K(x)  &= \sum_{i=1}^n c_i^K  x_i,
\end{split}
\end{align}
where $c_i^k \in \Z$ for $i=1,\ldots,n$ and $k=1,\ldots,K$.
Its name stands for
\emph{Unconstrained Multi-Objective Combinatorial Optimization} 
problem and we
will study it in more detail in our benchmark analysis. 
We note that to the problem above, 
we can canonically assign a decision problem and a counting problem.
The decision problem is of the form:
\begin{align*}
    \text{Given $b\in \Z^n$. Does there exist a $x \in \F[n]$ such that $f(x) \leqq b$?}
\end{align*}
The counting problem is of the form:
\begin{align*}
    \text{Given $b\in \Z^n$. How many $x \in \F[n]$ exist such that $f(x) \leqq b$?}
\end{align*}
It can be shown (i.e. by giving a polynomial-time reduction to the 
knapsack problem \cite{Ehr2005}) 
that even for $K=2$, the decision version of \eqref{eq:UMOCO_complexity} 
is NP-complete and the counting version is $\#$P-complete.
It follows that \eqref{eq:UMOCO_complexity} is $\#$P-hard (and therefore NP-hard).
Note that the single-objective problems, of which an \eqref{eq:UMOCO_complexity} 
instance is composed of, can be solved in polynomial-time,
thus the complexity arises from the multicriteria nature of the problem.

Given the inherent difficulty of the problem, heuristic strategies are a central focus of research. 
The next section outlines such an approach, which explores the use of quantum computing.
\section{Variational Quantum Algorithm for Multicriteria Optimization}\label{sec:variational}

In this section, we briefly describe a heuristic algorithm for solving 
\textbf{MCO2} with a hybrid classical-quantum solver.
More specifically, we consider a variational quantum algorithm
for solving binary multicriteria optimization problems, which was first introduced in \cite{Eks2024}. 
This algorithm will serve as the starting point for our studies.
We assume knowledge of the basic concepts of the mathematical formalism of quantum computing
\cite{Kit2002,Nie2010}
and the main concepts of variational algorithms \cite{Cer2021,Per2013,Til2022}. 
We will use the notation $\B := \C^2$ and 
$\Bn := \C^2 \otimes \cdots \otimes \C^2$
for the complex euclidiean vector spaces used for describing quantum computation.

Among the key steps in designing a variational quantum algorithm are the definition of an ansatz, 
the choice of a cost function, and the integration of a classical optimizer.
Motivated by the research surrounding QAOA
(a VQA for single-objective combinatorial optimization
\cite{Far2014}), 
a variant of QAOA for multicriteria combinatorial optimization problems 
was presented in \cite{Eks2024}. We provide a brief overview.

\subsection{Ansatz}
Let $f = (f_1, \ldots, f_K)$ be the objective functions of 
a multicriteria optimization problem, where $f_k: \F[n] \longrightarrow \R$
for $k=1,\ldots,K$. 
Each element of $\F[n]$ is identified with a 
computational basis state of $\Bn$.
More precisely, $x = (x_1,\ldots,x_n) \in \F[n]$ is encoded as 
$\ket{x}= \ket{x_1} \otimes \cdots \otimes \ket{x_n} \in \Bn$. 
To each of the objective functions $f_k$, we assign a diagonal operator 
$H_k$ on $\Bn$, which operates on the computational basis as follows,
\begin{align*}
    H_k\ket{x} = f_k(x) \ket{x}.
\end{align*}
It can be shown \cite{Luc2014} that for many functions $f_k$, the above operator 
can be implemented on a quantum computer efficiently (without explicitly 
calculating all $2^n$ values of $f_k$).
This applies, for example, to all linear and quadratic polynomials, 
in particular to the functions in our benchmark analysis. 
This hermitian operator $H_k$ induces a parameterized quantum gate,
\begin{align*}
    U_{k}(\gamma_k) := e^{-i \gamma_k H_k}, \quad \gamma_k \in \R.
\end{align*}

An additional parameterized quantum gate is introduced before the unitaries of different 
$f_k$ are combined; here, in contrast to the step above, 
only the variational parameter depends on the objective function, 
while the Hermitian operator remains fixed,
\begin{align*}
    U_M(\beta_k) = e^{-i \beta_k\sum_{j=1}^n X_j}.
\end{align*}
Each summand $X_j$ in the exponent is the hermitian operator defined by 
\begin{align*}
    X_j = I \otimes \cdots  \otimes I \otimes X \otimes I \otimes \cdots \otimes I,
\end{align*}
where $I$ is the identity on $\B$. 
Here, $X$ appears on the $j$-th 
factor of the tensor product and operates as $X\ket{0}=\ket{1}$ and 
$X\ket{1}=\ket{0}$. 
The operator $U_k(\gamma_k)$ is called \emph{problem unitary} (with respect
to $f_k$)
and $U_M(\beta_k)$ is called the \emph{mixer unitary}.
The $2K$ operators are combined into one layer by alternated multiplication,
\begin{align*}
    U_l =  U_M(\beta_{l,K})U_K(\gamma_{l,K}) \cdots U_M(\beta_{l,1})U_1(\gamma_{l,1}).
\end{align*}
We used the index $l \in \N$ 
to indicate the dependency of the parameters on the layer.
The finalized ansatz is created by applying different layers in succession,
\begin{align*}
    U = U_L \cdots U_1,
\end{align*}
with $L \in \N$.
The ansatz $U$ depends on a total of $L2K$ parameters. 
In most cases, it is not important to distinguish between $\gamma$ and $\beta$ parameters, 
so we summarize all parameters to a vector $\theta \in \R^{L2K}$,
\begin{align*}
    \theta = ( \gamma_{1,1}, \beta_{1,1}, \ldots, \gamma_{L,K}, \beta_{L,K}),
\end{align*}
and write $U = U(\theta)$.

\subsection{Cost Function}
First, we clarify how a subset of 
the decision space $\F[n]$ is derived from the quantum state generated by the ansatz $U(\theta)$.
The inital state $\ket{\varphi}$ is chosen as
\begin{align*}
    \ket{\varphi} := \bigotimes_{i=1}^n \frac{1}{\sqrt{2}}\left( \ket{0}+\ket{1}\right) \in \Bn
\end{align*}
and let $\ket{\varphi(\theta)}$ be the state generated by the ansatz,
\begin{align*}
    \ket{\varphi(\theta)} = U(\theta)\ket{\varphi}.
\end{align*}
As an element of the vector space $\Bn$, the state $\ket{\varphi(\theta)}$ is a linear combination of the vectors 
in the computational basis,
\begin{align*}
    \ket{\varphi(\theta)} = \sum_{x \in \F[n]} \alpha_x \ket{x},
\end{align*}
where $\alpha_x \in \C$ such that $\sum_{x} \abs{\alpha_x}^2 = 1$.
The idea, for constructing a candidate set of solutions $\X \subseteq \F[n]$,
is to consider those elements $x \in \F[n]$, for which $\ket{x}$ 
has a large coefficient.
In practice, the absolute values of the coefficients $\alpha_i$ are estimated by repeated measurement.
The number of solutions we want to extract from the quantum state 
(which is therefore equal to the number of coefficients to be considered) is denoted 
as $P$ (number of Pareto points).
The parameter $P \in \N$ is therefore a hyperparameter of the algorithm.
The subset of the decision space containing the most frequent basis elements in the above
linear combination is denoted as $\X(\theta)$,
\begin{align*}
    \X(\theta) := \Menge{x \in \F[n]}{\text{$\ket{x}$ is among the 
    $P$ most frequent basis vectors in $\ket{\varphi(\theta)}$}}.
\end{align*} 
The corresponding image in objective space is denoted as 
\begin{align*}
    \Y(\theta) :=f(\X(\theta)).
\end{align*}
To measure the quality of $\Y(\theta)$, the hypervolume 
$\HV(\Y(\theta))$
is calculated.
Therefore, the cost function to be minimized is defined as 
\begin{align*}
    c: \R^{L2K} \longrightarrow \R, \quad \theta \longmapsto -\HV(\Y(\theta)).
\end{align*}

The variational quantum algorithm that uses the above ansatz and the above cost function
is referred to as QMOO in the rest of the paper.

\section{Combining Algorithms from Multicriteria Metaheuristics with Variational
Quantum Algorithms} \label{sec:qnsgaii}

In the context of variational quantum algorithms, and in particular for the algorithm 
from \Cref{sec:variational}, 
usually standard algorithms for single-objective optimization
(such as Nelder-Mead, Powell, BFGS, etc.) are used to minimize the cost function.
This follows from the design decision, in which the search for good variational parameters is translated, 
through the use of a cost function, into a real-valued optimization problem.

In this section, we propose two new formulations of the variational 
quantum algorithm from \Cref{sec:variational}.
One formulation in which classical algorithms from various multicriteria-metaheuristics can be 
used to find suitable parameters,
and another that is specifically designed for the application of NSGA-II 
as a classic solver.

In essence, when the state $\ket{\varphi(\theta)}$ generated by the variational ansatz, is translated
into a subset $\Y(\theta) \subseteq \R^K$ of the objective space, 
the set is not evaluated using the hypervolume indicator,
but instead, the set-valued function $\theta \longmapsto \Y(\theta)$ is interpreted as a 
multicriteria optimization problem, 
which is accessible for classical metaheuristics.

The technical details are discussed below, but first we would like to 
mention the two main reasons why this approach can be advantageous.
First, 
when translating a multicriteria problem into a single-criteria problem (as usually done 
when applying cost-functions), 
some of the information about the Pareto front is usually lost. 
By not significantly changing the nature of the optimization problem, we expect this effect to be reduced.
Second, our proposed formulation enables the use of a wide range of metaheuristics, 
which could prove useful in combination with quantum computing. 
Notably, population-based algorithms applied to the variational 
ansatz generate several Pareto front approximations simultaneously.

We will start with the general approach and 
then discuss adjustments specifically tailored to NSGA-II.

\subsection{General Approach}
Let $f = (f_1, \ldots, f_K)$ be the objective functions of 
a multicriteria optimization problem, where $f_k: \F[n] \longrightarrow \R$
for $k=1,\ldots,K$. 
To formulate the above idea precisely, let 
\begin{align*}
    \PF := \Menge{\X \subseteq \F[n]}{\# \X = P}
\end{align*}
be the set of all subsets of $\F[n]$ with cardinality $P$.
We recall that the bitstrings in $\F[n]$ can be identified in a natural way with integers, 
so that one always has a natural ordering of elements in $\X \subseteq \F[n]$.

Let $\X^1, \X^2 \in \PF$ be given, where the indexing of its elements corresponds to the natural ordering of integers,
\begin{align*}
    \X^1 = \menge{x_1^1, \ldots, x_{P}^1}, \quad \X^2 = \menge{x_1^2, \ldots, x_{P}^2}.
\end{align*}
Let $\Sym$ be the symmetric group on $P$ elements.
We define a partial order on $\PF$ by specifying that the set $\X^1$ dominates the set $\X^2$, 
if the elements in the two sets can be arranged such that 
the relation $\preceq_f$ (see \Cref{sec:mco}) holds elementwise,
\begin{align*}
    \X^1 \trianglelefteq_f \X^2 : \iff \exists\, \sigma \in \Sym: x_i^1 \preceq_f x_{\sigma(i)}^2\, \forall i = 1,\ldots, P.
\end{align*}
This relation is indeed a partial order. Reflexivity follows by taking the identity permutation, 
transitivity follows by taking the product of two permutations, 
and antisymmetry follows from the fact that each element of $\Sym$ has finite order.

So given our initial objectives $f=(f_1,\ldots,f_K)$, we can associate to it a new optimization problem as follows.
\begin{description}
    \item[(MCO3)] \text{Determine $\min \PF$ with respect to $\trianglelefteq_f$.}
\end{description}
The connection between $\min \PF$ and the Pareto front $\min f(\F[n])$ is 
as follows:
\begin{enumerate}
\item If $P = \# \min f(\F[n])$,
i.e. if $P$ is equal to the number of points on the Pareto front,
then the set (of subsets) $f(\min \PF)$ consists of one element. 
This element is equal to the Pareto front.
In this case, solving \textbf{MCO1} can be reduced to solving \textbf{MCO3}.
\item If $P < \# \min f(\F[n])$, then every element in $f(\min \PF)$ will be a subset of the Pareto front.
\item If $P > \# \min f(\F[n])$, then every element in $f(\min \PF)$ will contain a subset of the Pareto front.
\end{enumerate}
In the last two cases, solving \textbf{MCO3} usually produces good candidates for \textbf{MCO2}.

Next, we show how to adapt QMOO
to obtain a VQA for \textbf{MCO3}.
The idea is to define a function $F: \R^{L2K} \longrightarrow \R^{KP}$ using the variational ansatz
$U(\theta)$, 
such that the associated multicriteria optimization problem with respect to $\preceq_F$ 
has the "same" minimal elements as \textbf{MCO3}. 
This yields a hybrid classical-quantum algorithm 
in which the function $F$ to be optimized
can be evaluated using a quantum computer, 
but the classical optimizer can be any algorithm for multicriteria optimization.

Let $U(\theta)$ be our variational ansatz, let $\ket{\varphi}$ be some initial state 
and let $\ket{\varphi(\theta)}$ the state generated by the ansatz,
\begin{align*}
    \ket{\varphi(\theta)} = U(\theta)\ket{\varphi}.
\end{align*}
As discussed in \Cref{sec:variational}, the state $\ket{\varphi(\theta)}$ 
induces a set $\X(\theta)$ of most frequent elements in the standard base representation of $\ket{\varphi(\theta)}$.
Let the elements of $\X(\theta)$ be sorted with respect to the natural ordering of $\Z$,
\begin{align*}
    \X(\theta) = \menge{x_1, \ldots, x_{P}}.
\end{align*}
Let $\sigma$ be the permutation that sorts the elements of $\X(\theta)$ 
according to their frequency in the base representation of $\ket{\varphi(\theta)}$.
That is, $x_{\sigma(1)}$ has the largest coefficient in  $\ket{\varphi(\theta)}$, then
$x_{\sigma(2)}$  the second largest, and so on.
The set $\X(\theta)$ is transformed into a vector from $\R^{KP}$ as follows,
\begin{align*}
    \vc_f({\X(\theta)}) := (f_1( x_{\sigma(1)} ), \ldots, f_K( x_{\sigma(1)}), 
    \ldots, 
    f_1( x_{\sigma(P)} ), \ldots, f_K( x_{\sigma(P)})).
\end{align*} 
Overall, we can define the following map using a variational ansatz $U(\theta)$,
\begin{align*}
    F: \R^{L2K} \longrightarrow \R^{KP}, \quad \theta \longmapsto \vc(\X(\theta)).
\end{align*}
The motivation for this construnction is based on the following observation,
\begin{align*}
    \text{
        $\X(\theta)$ is minimal in $\PF$ w.r.t. $\trianglelefteq_{f} 
        \iff 
        F(\theta)$ is minimal in $F(\R^{KP})$ w.r.t $\leqq$.
    }
\end{align*}
We have reduced \textbf{MCO3} to an (ordinary) multicriteria optimization problem 
with respect to the 
(ordinary) Pareto dominance relation in the sense of \Cref{sec:mco}.
The resulting benefit is that the problem on the right-hand side can be 
approached by a variational quantum algorithm, 
whose variational parameter can be searched using algorithms from any classical metaheuristic.
Since solving \textbf{MCO3} produces good candidates for \textbf{MCO1} and \textbf{MCO2}, 
the right hand side above defines a hybrid quantum-classical algorithm for 
our inital problem of interest.

In the case of NSGA-II, the above idea can be further extended to 
adapt the special characteristics of NSGA-II 
(namely the interaction of non-dominated sorting and crowding distance) 
to the variational quantum algorithm.

\subsection{NSGA-II as Classical Solver}
The Nondominated Sorting Genetic Algorithm II (NSGA-II), 
introduced by Deb et al. in 2002~\cite{nsga2}, 
is one of the most widely used evolutionary algorithms for 
solving multicriteria optimization problems. 
Before we utilize our framework to apply NSGA-II for 
the variational parameter optimization, 
we shortly introduce the components of NSGSA-II.

\begin{description}
	\item[Population and Evaluation] The population consists of potential solutions to the problem
	\item[Nondominated Sorting] NSGA-II uses a ranking mechanism to classify solutions based on the dominance relation. The population is sorted into a hierarchy of fronts, where the first front contains all efficient solutions, the Pareto front, the second front contains solutions dominated only by those in the first front, and so on. This sorting allows the algorithm to prioritize solutions closer to the Pareto front in the selection process.
	\item[Crowding Distance] To maintain diversity among the solutions, NSGA-II introduces the crowding distance metric. For each solution within a Pareto front, the crowding distance measures the density of solutions surrounding it in the objective space. This is computed as the perimeter of the hyperrectangle formed by the nearest neighbors in each objective dimension. Solutions with a larger crowding distance are preferred during selection, ensuring that the algorithm explores diverse regions of the Pareto front.
	\item[Elitist Selection and Genetic Components] The algorithm employs an elitist strategy to combine the parent and offspring populations and selects the best individuals for the next generation. The selection is based on Pareto rank and crowding distance. Specifically, solutions are compared using a binary tournament selection, which randomly selects individuals and prefers lower-rank individuals and breaks ties by favoring those with higher crowding distance. This elitism ensures that high-quality solutions are chosen to produce the next offspring. Specifically, the offspring is generated from these parents by employing specific crossover and mutation rules.
\end{description}

This is applied iteratively until a certain termination criteria 
(for example maximum number of generations) is met.

\begin{figure}[tb]
	\includegraphics[width=0.5\textwidth]{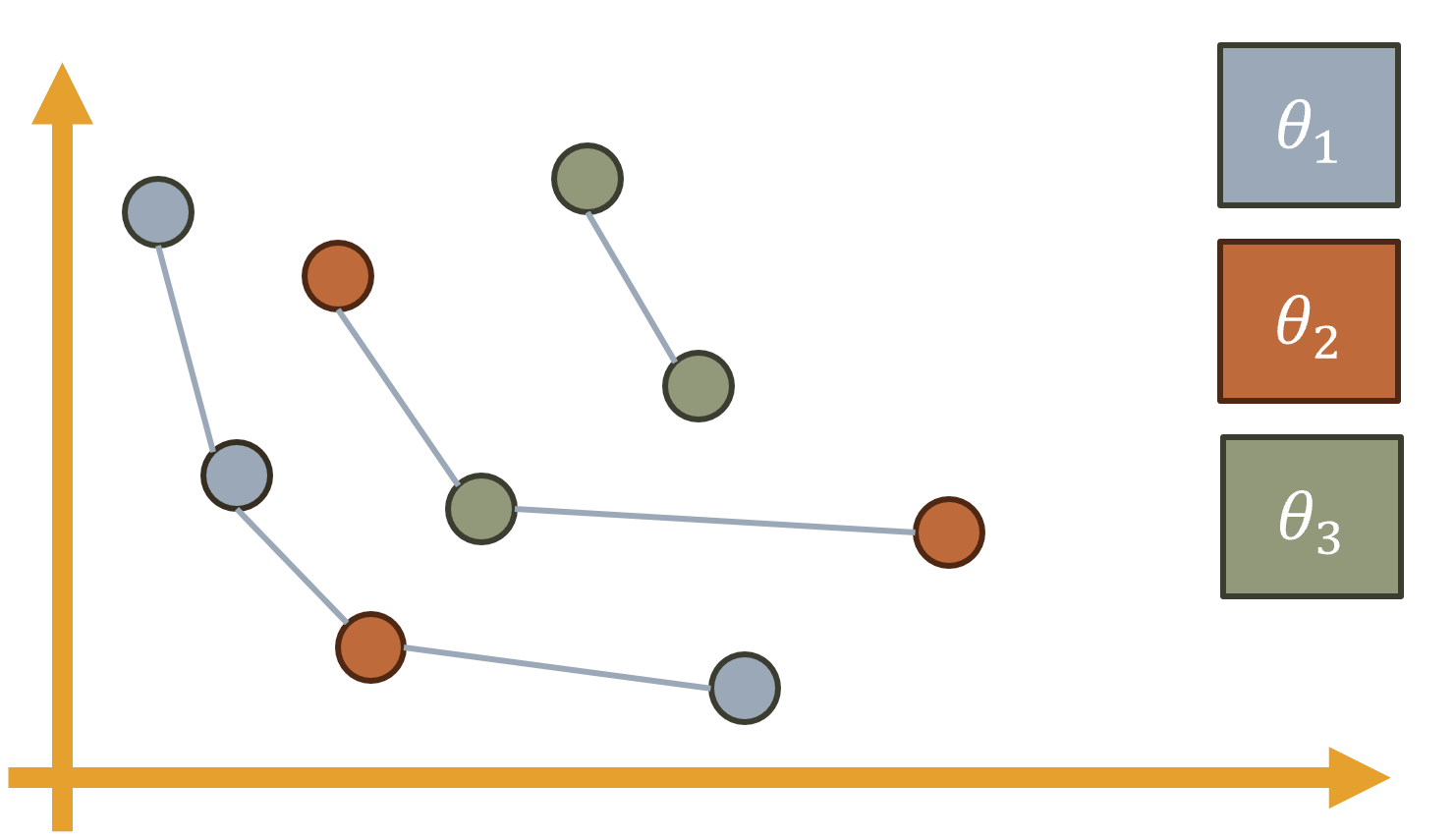}
	\caption{An example execution of the nondominated sorting: We have three variational parameter each having three elements. We perform nondominated sorting for the union of all elements and rank them accordingly. Now, we see that $\theta_1$ and $\theta_2$ both have an element in the first Pareto front, thus, these parameters are ranked accordingly. Parameter $\theta_3$'s best element has a rank of 2, thus $\theta_3$ gets rank 2.}\label{fig:nondomsorting}
\end{figure}

With that, we can adjust NSGA-II to optimize the variational parameters of QMOO as follows: NSGA-II is initialized with a population of variational parameters $\theta$. The corresponding quantum circuits are executed and the most frequent elements $X(\theta)$ are observed. Now, for both nondominated sorting and the calculation of the crowding distance, we consider the union $\mathfrak{X}\coloneqq \bigcup_{\theta\in Pop} X(\theta)$. We apply the standard nondominated sorting and the crowding distance to $\mathfrak{X}$ and its elements and assign the results to the elements. Then, the rank of an individual $\theta$ is determined by the minimal rank among its elements in $X(\theta)$ and its crowding distance of the best crowding distance of its elements. Figure~\ref{fig:nondomsorting} shows an example for the nondominated sorting.

This algorithm has been implemented in Python using the pymoo 
library~\cite{pymoo}, where we have modified the existing functions to match our approach. Specifically, we have used the following options and parameters that, in tests before conduction the computational study have turned out to be superior:
\begin{itemize}
	\item The population size (number of variational parameters) is set to 5. It has been initialized using pymoo's FloatRandomSampling. For the execution of a quantum circuit the sample set has size of the sum of the number of variables and objective functions.
	\item The nondominated sorting and crowding distance are the standard ones of NSGA-II in pymoo (modified to our problem).
	\item As tournament selection, we use binary tournament selection.
	\item Crossover is executed using simulated binary crossover with eta=15 and a probability of 90\%.
	\item For the mutation, we use polynomial mutation with eta=20.
	\item As a termination criteria, we use the standard values of pymoo's DefaultMultiobjectiveTermination criteria and just set the maximal number of generations to 200 and the maximum number of evaluations to 4000. 
\end{itemize}

The variational quantum algorithm that uses the ansatz $U(\theta)$ and the above
cost function is referred to as QMOOM in the rest of the paper.
We demonstrate this approach in our benchmark analysis by using NSGA-II as a classical optimizer.
\section{Increasing the Coverage of the Pareto front Approximation}\label{sec:spread}
As mentioned in the first section, the hypervolume indicator is a well-established measure
for evaluating the quality of a Pareto front approximation.
It is often used in algorithms to guide iterative approximation procedures 
and to aid
convergence towards the true Pareto front \cite{Zit1998, Kno1999,Emm2006}.
This makes the hypervolume a natural first choice in the design of
a variational quantum algorithm for multiobjective optimization.

\begin{figure}[htbp]
    \includegraphics[scale=0.4]{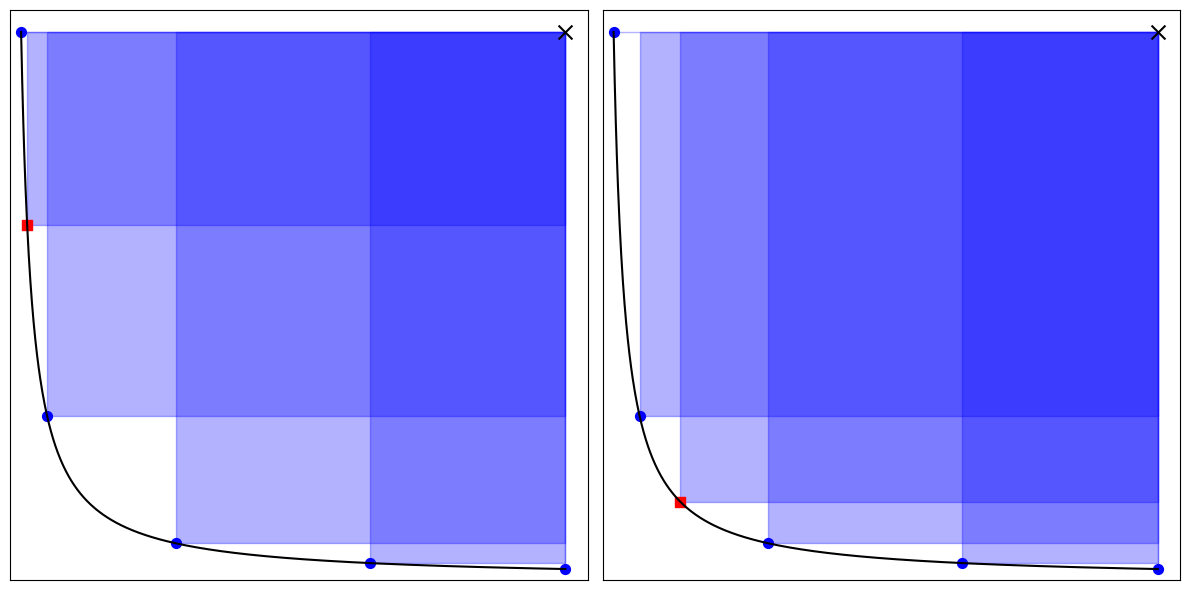}
    \centering
    \caption{An example that demonstrates how an increase in hypervolume can be accompanied by a reduction in coverage.
    The black curve is the Pareto front, the points on the curve are the Pareto front approximation.
    In the left figure, the points are evenly distributed.
    In the right figure, the red point (marked as square) is shifted to a new position,
    thereby increasing the hypervolume but reducing the uniformity.}
    \label{fig:coverage_decreased}
\end{figure}

Despite its favorable convergence properties, 
a high hypervolume may correspond to a Pareto front approximation
that fails to adequately "cover" the true Pareto front.
Below, we will define in more detail various notions of measuring "coverage", but
intuitively, good coverage means that points in the approximation are evenly distributed
and that “all areas" of the Pareto front are represented.
\Cref{fig:coverage_decreased} illustrates a simple example in which the 
hypervolume increases as the coverage of the approximation decreases.

High coverage is, of course, motivated by practical requirements.
It ensures that the decision maker has a large number of compromises
with different tradeoffs.
It allows to make an informed decision based on a
complete overview of possible solutions, rather than relying only on certain clusters,
where potentially favorable decisions could be overlooked.

Theoretically, a direct connection between hypervolume and coverage can only be
established for simple examples.
For bicriteria problems, maximizing hypervolume is equivalent to maximizing 
coverage 
only when the Pareto front forms a straight line.
However, in most cases, the distribution of points, which maximize the hypervolume,
depends significantly on the geometric
properties of the Pareto front, such as its curvature,
and for more complex objective functions,
there is no theoretical guarantee that maximizing hypervolume will also yield a high coverage
\cite{Aug2009}.
Specifically, QMOO is not expected to achieve an even distribution
along the true Pareto front.

In the following, we examine whether incorporating additional coverage measures alongside hypervolume can
enhance the performance of the variational algorithm.
Since coverage indicators by themselves do not drive convergence to the true Pareto front,
we take the average of a coverage indicator and the hypervolume.

To define the notation, let $\I(\Y(\theta))$ be some indicator
evaluated on $\Y(\theta)$
and $\HV(\Y(\theta))$
the hypervolume.
We will consider the ansatz from \Cref{sec:variational} with the
cost function
\begin{align*}
    p I(\Y(\theta)) + (1-p)(-\HV(\Y(\theta))),
\end{align*}
where $p \in [0,1]$.
We choose this average-based cost function over an 
indicator that jointly promotes convergence and coverage,
as it facilitates a clearer evaluation of the trade-off 
between the two and allows us to quantify the extent 
to which coverage can be prioritized without significantly 
compromising convergence.

The indicator $\I$ and the weight $p$ are therefore hyperparameters of the algorithm.
In our experiments in the next section, 
we use a total of six different indicators for $I$ and test their 
performance.
The variational algorithm consisting
of the ansatz $U(\theta)$ from \Cref{sec:variational} and the above cost-function,
will henceforth be referred to as $\mathrm{QMOOC}(\I,p)$.

In the following, we collect all the concepts needed to define the indicators for our 
experimental study.
For a overview of various indicators, please refer to \cite{Aud2020}.

Let $f = (f_1,\ldots, f_K)$ be our objective with
$f_k:\F[n] \longrightarrow \R$ for $k=1,\ldots,K$.
Let $\X \subseteq \F[n]$ be an arbitrary subset of the decision space
and $\Y=f(\X) \subseteq \R^K$
the corresponding subset of the objective space.

Let $x^{(k)}$ be a minimum of the single-criteria problem corresponding to $f_k$,
and $y^{(k)}:=f(x^{(k)})$.

For $z \in \R^K$, we say
\begin{align*}
    \nb(z) := \argmin_{y \in \Y \setminus \{z\}} \norm[2]{y-z} \in \Y
\end{align*}
is the \emph{closest neighbor} of $z$ in $\Y$ (with respect to the euclidiean norm).

We fix an arbitrary enumeration of the elements in $\Y$,
\begin{align*}
    \Y = \menge{y^1, y^2, \ldots, y^{\abs{\Y}}}.
\end{align*}
For each objective function $f_k$, let $\sigma_k \in \mathrm{Sym}(\abs{\Y})$
be the permutation that sorts the elements in $\Y$
according to their $k$-th coordinate in ascending order,
\begin{align*}
    y_k^{\sigma_k(1)} \leq y_k^{\sigma_k(2)} \leq \ldots \leq y_k^{\sigma_k(\abs{\Y})}.
\end{align*}
Ties are broken arbitrarily.
With this, we define for $k \in \menge{1,\ldots,K}$ and $j \in \menge{1, \ldots,\abs{\Y}-1}$,
\begin{align*}
    d_{k,j} := y_k^{\sigma_k(j+1)}-y_k^{\sigma_k(j)}.
\end{align*}

The selection of indicators below is motivated by the desire to capture
various important “coverage properties”, such as the extent of the approximation,
the distribution of points, its uniformity, and boundary coverage.

Introduced in \cite{overallparetospread}, the  
\emph{Pareto Spread} captures the extent of the Pareto front covered by the approximation,
\begin{align*}
    \PS(\Y) = \prod_{k=1}^K 
    \frac{
        \abs{\max_{y \in \Y}y_k - \min_{y \in \Y}y_k}
        }{
            \abs{\tilde{y}_k^{I} - \tilde{y}_k^{N}}
            },
\end{align*}
where $\tilde{y}^I$ is an approximation of the ideal and $\tilde{y}^N$ an approximation of
the nadir.
It computes the ratio of the range of objective values covered by the approximation, 
relative to the (approximations of the) ideal and nadir. 
It provides insight into the comprehensiveness of an approximation, 
although it does not directly account for the distribution of points along the front.

The \emph{Outer Diameter} indicator, introduced by \cite{Ziz2008}, 
assesses the maximum extent of a Pareto front approximation along its objective dimensions, 
\begin{align*}
    \OD(\Y) &:=  \max_{k=1,\ldots,K} \max_{y,\tilde{y} \in \Y} \abs{y_k - \tilde{y}_k}.
\end{align*}
Despite being computationally efficient, it only evaluates the largest distance along 
a single dimension, potentially neglecting overall spread information.

The \emph{$\M$-indicator}, originally proposed by \cite{Ziz2000}, 
is defined as,
\begin{align*}
    \M(\Y) &:= \sqrt{\sum_{k=1}^{K} \max_{y,\tilde{y} \in \Y} \abs{y_k - \tilde{y}_k}},
\end{align*}
It particularly measures how well the approximation covers the extremes of the Pareto front, 
serving as a robust metric for assessing boundary coverage and ensuring comprehensive 
optimization across objective extremes

Proposed by \cite{Zhe2017}, the \emph{Distribution Metric} corrects the 
limitations of above mentioned measures by normalizing distances between solutions and 
addressing discrepancies due to varying objective magnitudes. 
It evaluates both distribution uniformity and the extent of the Pareto front, 
providing a balanced view of solution diversity and coverage quality.
For each objective $f_k$, consider mean and standard deviation over the $d_{k,j}$,
\begin{align*}
    m_k &= \frac{1}{\abs{\Y}-1} \sum_{j=1}^{\abs{\Y}-1} d_{k,j}, \\
    s_k &= \frac{1}{\abs{\Y}-2} \sum_{j=1}^{\abs{\Y}-1} \left( d_{k,j} - m_k\right)^2.
\end{align*}
With the ideal $y^I$ and nadir $y^N$, the distribution metric is defined as
\begin{align*}
    \DM(\Y):=\frac{1}{\abs{\Y}} \sum_{k=1}^K \frac{s_k}{m_k}
    \left( \frac{\abs{y_k^I - y_k^N}}{\max_{y \in \Y} y_k - \min_{y \in \Y} y_k} \right).
\end{align*}

Introduced in \cite{Zho2006}, the \emph{$\D$-indicator} evaluates the spread of Pareto front 
approximations by analyzing the distribution of solutions across the objective space. 
It measures variations in spacing between adjacent solutions and incorporates the overall 
extent of the Pareto front approximation. 
This provides an integrated assessment that accounts for both the uniformity of solution 
spacing and the boundary coverage It is defined as follows,
\begin{align*}
    \D(\Y) := \frac{
        \sum_{k=1}^{K} \norm[2]{y^{(k)}-\nb(y^{(k)})} + \sum_{y \in \Y} \abs{\mu_2-\norm[2]{y - \nb(y)}}
    }{
        \sum_{k=1}^{K} \norm[2]{y^{(k)}-\nb(y^{(k)})} + \abs{\Y} \cdot \mu_2
    },
\end{align*}
where
\begin{align*}
    \mu_2 := \frac{1}{\abs{\Y}} \sum_{y \in \Y} \norm[2]{y-\nb(y)}
\end{align*}
is the mean.

The \emph{Evenness indicator} by Messac \cite{evenness} 
specifically quantifies how uniformly distributed solutions are along the Pareto front 
approximation. 
It provides a straightforward measure of distribution quality without dependence on 
external parameters, 
\begin{align*}
    \EV(\Y) :=
    \frac{
        \max_{y \in \Y} \min_{y' \in \Y \setminus \menge{y}} \norm[2]{y-y'}
    }{
        \min_{y \in \Y} \min_{y' \in \Y \setminus \menge{y}} \norm[2]{y-y'}
    }.
\end{align*}

\section{Experiments}\label{sec:experiments}
The modifications to the variational quantum algorithm proposed in 
\Cref{sec:qnsgaii} and \Cref{sec:spread}
were tested on several benchmark functions.
We consider unconstrained bicriteria optimization problems 
$f = (f_1, f_2)$
whose objective functions $f_1, f_2: \F[n] \longrightarrow \R$ are both either linear 
or quadratic polynomials in $n$ variables. 
In total, we consider four types of benchmark problems, two linear and 
two quadratic types.
To facilitate comparability with \cite{Eks2024}, 
three of the four types 
are constructed in the same way as in \cite{Eks2024}.
In the following, we will discuss the benchmark functions in more detail.
\Cref{tab:overview_benchmark_functions} gives a brief overview.
\begin{table}[htbp]           
    \centering                  
    \begin{tabular}{lr}
      \toprule              
      Linear types & UMOCO-1, UMOCO-2 \\
      Quadratic types & AFM, FM-AFM \\
      Number of objectives & $2$ \\
      Number of variables & $10$ and $13$ \\
      Number of instances per problem type & $20$ \\
      \bottomrule
    \end{tabular}
    \caption{Overview of our benchmark problems.}
    \label{tab:overview_benchmark_functions}
\end{table}

Let $U(a,b)$ denote the uniform distribution on the intervall $(a,b)$ and
let $U\{a,b\}$ the uniform distribution on integers in $(a,b)$, where $a,b \in \R$.
The first two problem types are variants of the UMOCO problem 
mentioned in \Cref{sec:mco},
\begin{align*}
    f_1(x)  &= \sum_{i=1}^n c_i^1  x_i, \\
    f_2(x)  &= \sum_{i=1}^n c_i^2  x_i.
\end{align*}
The difference between the two variants is the way in which the 
coefficients are sampled, 
which also results in differences in properties of the Pareto front.

In the first type, UMOCO-1, the coefficients $c_i^1$ are sampled from $U(-1,1)$.
To ensure the two objectives are conflicting, the 
coefficients $c_i^2$ are defined as 
\begin{align*}
    c_i^2 := -\tfrac{1}{2}c_i^1 + \tfrac{1}{2}d_i,
\end{align*}
where $d_i$ is sampled from $U(-1,1)$.

In the second type, UMOCO-2, 
the coefficients are randomly sampled according to $U\{-10n,10n\}$.
A tradeoff between the two objective functions is enforced by
only considering instances, where the coefficient-vectors
$(c_1^1, \ldots, c_n^1)$ and 
$(c_1^2, \ldots, c_n^2)$ form an angle between $90$ and $150$ degrees.
The objective functions are then scaled to have the same order of 
magnitude as UMOCO-1.

Next, we consider benchmark problems that are given by quadratic functions of the form
\begin{align*}
    f_1(x)  &= x^{\intercal} J^1 x + x^{\intercal} h^1 + c^1, \\
    f_2(x)  &= x^{\intercal} J^2 x + x^{\intercal} h^2 + c^2,
\end{align*}
where $J^1, J^2 \in \R^{n \times n}$ are symmetric, $h^1, h^2 \in \R^n$ and $c^1, c^2 \in \R$.
We look at two types of quadratic problems, 
which again differ in the way the above coefficients are sampled.
Both variants are inspired by physical spin models, which lie beyond the scope of this paper.
We provide a brief summary of the construction.
For more details, please refer to \cite{Eks2024} and the github repository \cite{Sch2024}.

The first class of quadratic problems is abbreviated as FM-AFM 
(standing for FerroMagnetic and 
Anti-FerroMagnetic). 
Here, the coefficients of $J^1$ are sampled with respect to $U(a,b)$ with 
$a < b < 0$, and the coefficients of $J^2$ are sampled with respect to $U(a,b)$ with 
$b > a > 0$. For $i=1,2$, the vector $h^i$ is constructed as 
\begin{align*}
    h^i = g^i - \tfrac{1}{2} \mathds{1}^{\intercal} J^i,
\end{align*}
where $g^i \in \R^n$ is a vector, whose entries are sampled according to $U(-1,1)$ and $\mathds{1}$ 
is the vector with all entries equal to $1$. The constants $c_1, c_2 \in \R$ are arbitrary.

The second class of quadratic problems is calld AFM. 
Here, $f_2$ is construced as above. Regarding $f_1$, the matrix $J^1$ is diagonal, where
the diagonal elements are sampled according to $U(a,b)$ with $0 < a < b$. The vector $h^1$ is defined 
as 
\begin{align*}
    h^1 = -2v^{\intercal}J^1,
\end{align*}
where the entries of $v \in \R^n$ are sampled according to $U(0,1)$.
Again, the constants $c_1, c_2 \in \R$ are arbitrary.

Each of the four problem types was considered with $n = 10$ and $n = 13$ variables 
This is due to hardware limitations when simulating quantum algorithms.
Nevertheless,  a solution space consisting of $2^{10}$ or $2^{13}$ 
elements can be considered a good representative.

For each of these $4\cdot2$ variants, $20$ random problem instances were generated. 
Each algorithm we examined was tested on all $4\cdot 2 \cdot 20$ problem 
instances, 
with each problem instance being executed on $40$ different seeds.

\begin{figure}[ht]
    \includegraphics[scale=0.124]{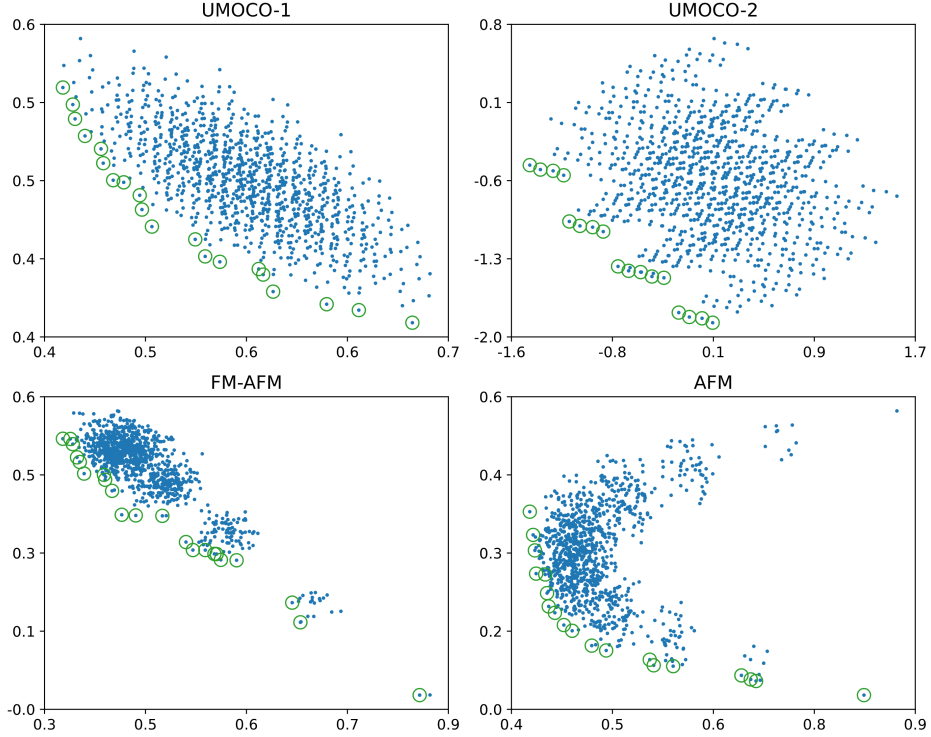}
    \centering
    \caption{For each of our problem types, the image $f(\F[n])$ 
    of a problem instance $f$ is shown as an example. 
    The points circled in green are the nondominated elements.}
    \label{fig:objective_spaces}
\end{figure}

\Cref{fig:objective_spaces} shows what a typical problem instance looks 
like for each problem type.

\subsection{Comparing QMOOM and QMOO}
For evaluation, we apply both algorithms to compute a Pareto front approximation, 
determine the resulting hypervolume, and compare it to the hypervolume of the true Pareto front
(which is the the largest possible hypervolume of all subsets of $f(\F[n])$).

Both, QMOOM and QMOO use the same variational ansatz $U(\theta)$
as described in \Cref{sec:variational}. 
We used $L=5$ layers in $U(\theta)$, which 
was a compromise between 
sufficient expressivity of the ansatz and the 
computational overhead when simulating the quantum circuit.
The inital parameters $\theta$ were chosen at random by sampling 
each component $\theta_i$ of $\theta$ according to $U(0,2\pi)$.

As described in \Cref{sec:variational}, QMOO utilizes the hypervolume as its cost function,
which was optimized using standard methods for real-valued functions. 
More specifically, we tested COBYLA \cite{Pow1994}, Nelder-Mead \cite{Nel1965}
and Powell \cite{Pow1964} as implemented in SciPy \cite{scipy}.

In contrast, QMOOM uses the ansatz $U(\theta)$ to define a vector-valued function, 
which we interpret as the objective function of a multicriteria optimization problem.
\Cref{tab:qmoo_qmoo-m} gives an overview.

The number of layers in the approach is limited to $5$, 
which is again due to hardware limitations.
We chose the number of solutions $P$ 
depending on the number of variables in order to accommodate 
the slightly different sizes in the Pareto front,
$P = \text{number of variables $+$ number of objectives}$.

The implementation and simulation of the quantum circuits was carried 
out using the Qiskit framework (version $1.3$) \cite{qiskit2024}.
In the case of $n=13$, we observed significant runtime problems when 
optimizing the circuit parameters with Nelder-Mead and Powell, 
so we limited ourselves to the “COBYLA” solver.

\begin{table}[htbp]           
    \centering                  
    \begin{tabular}{l|cc}
      \toprule              
       & QMOO & QMOOM \\
      \midrule
      Ansatz & $U(\theta)$ & $U(\theta)$ \\
      Number of layers $L$ & $5$ & $5$ \\
      Initial parameters & sampled via $U(0,2\pi)$ & sampled via $U(0,2\pi)$ \\
      Number of solutions $P$ & 12, 15 & 12, 15 \\
      Classical optimizer & \shortstack{COBYLA, Nelder-Mead, \\Powell} & NSGA-II \\
      \bottomrule
    \end{tabular}
    \caption{Overview of the setup used in comparing QMOO and QMOOM.}
    \label{tab:qmoo_qmoo-m}
\end{table}

\begin{figure}[ht]
  \includegraphics[scale=0.124]{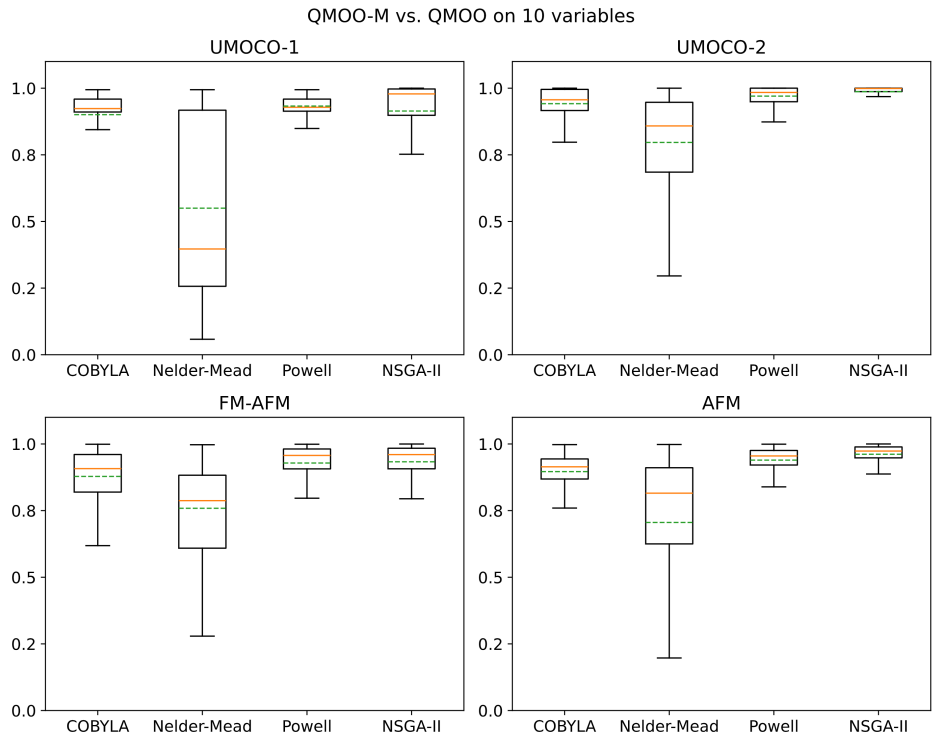}
  \centering
  \caption{The hypervolume achieved by QMOO and QMOOM relative to the hypervolume of the true Pareto front.
  The box plots are based on $800$ runs, considering $2$0 problem instances with $40$
  different seeds each. QMOO was combined with the classic solvers COBYLA, Nelder-Mead, 
  and Powell, while QMOOM was applied with NSGA-II. Here, problem instances with $n=10$ 
  variables were considered. 
  QMOOM consistently demonstrates superior 
  average performance, greater stability, and improved worst-case behavior.}
  \label{fig:qmoo-m_vs_qmoo_10}
\end{figure}

\begin{figure}[ht]
  \includegraphics[scale=0.124]{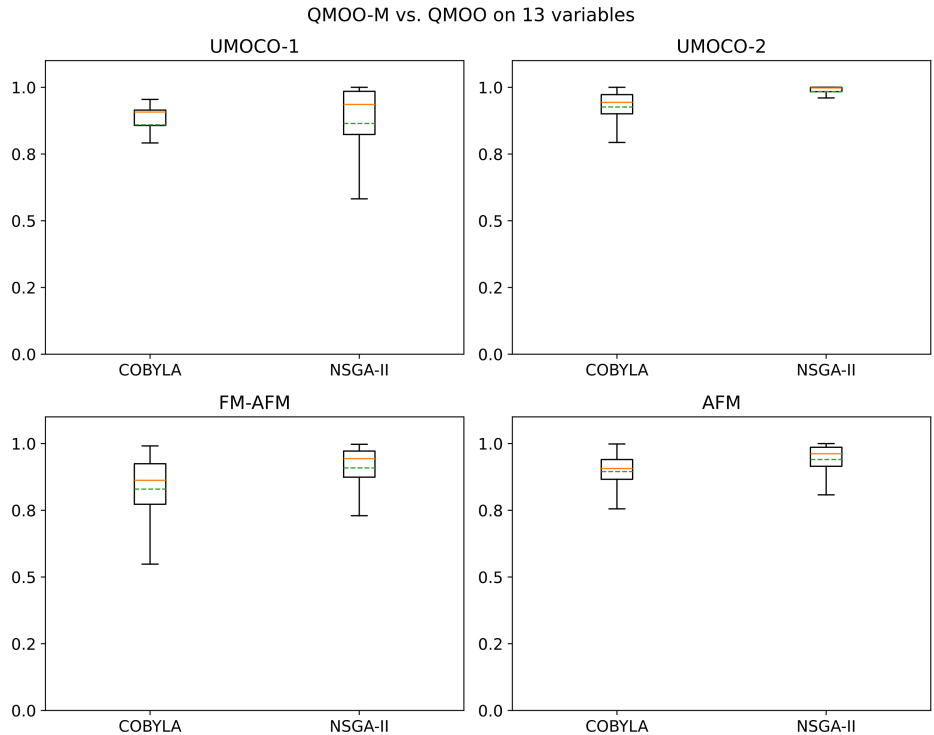}
  \centering
  \caption{The figure shows the situation analogous to \Cref{fig:qmoo-m_vs_qmoo_10} for $n=13$.
  Here, compared to the case $n=10$, the increase in average performance is slightly higher.}
  \label{fig:qmoo-m_vs_qmoo_13}
\end{figure}

\Cref{fig:qmoo-m_vs_qmoo_10} and \Cref{fig:qmoo-m_vs_qmoo_13} 
show the hypervolume achieved by QMOOM and QMOO relative to the 
hypervolume of the true Pareto front, when executed multiple times on instances of a particular 
problem type.

Each box plot is based on $800$ data points, 
which are generated by calculating $20$ problem instances, each $40$ times with different seeds.
The dotted green line marks the mean, the solid orange line marks the median.

One of the key observations is that QMOOM consistently achieves higher 
average performance across nearly all settings. 
The sole exception is a marginal dip in one experiment, specifically for 
$(n, \text{problem}, \text{solver}) = (10, \text{UMOCO-1}, \text{Powell})$, 
where its mean is only slightly lower. 
Depending on the solver used within QMOO, 
the improvement in mean performance spans from $1$ to $10$ percentage points, 
and occasionally even exceeds that, particularly when compared to Nelder-Mead.

Also noteworthy is QMOOMs stability; apart from UMOCO-1, 
the box plots are suggesting more consistent results across trials. 
Its advantage extends to the lower quantiles as well, 
again with UMOCO-1 being the only outlier, indicating that QMOOM offers
better worst-case performance in most cases.

The performance of both algorithms is more volatile in the case of $n=13$, 
but the increase in the average performance is here even more significant.

Overall, these results highlight QMOOMs consistent advantage 
in both average and worst-case performance, 
underscoring its robustness and effectiveness across a range of problem settings.

\subsection{Comparing QMOOC and QMOO}
We compare QMOOC and QMOO by analyzing the change in 
hypervolume and coverage when assigning a higher weight to the corresponding coverage 
indicator in the cost function.

\begin{figure}[ht]
  \includegraphics[scale=0.124]{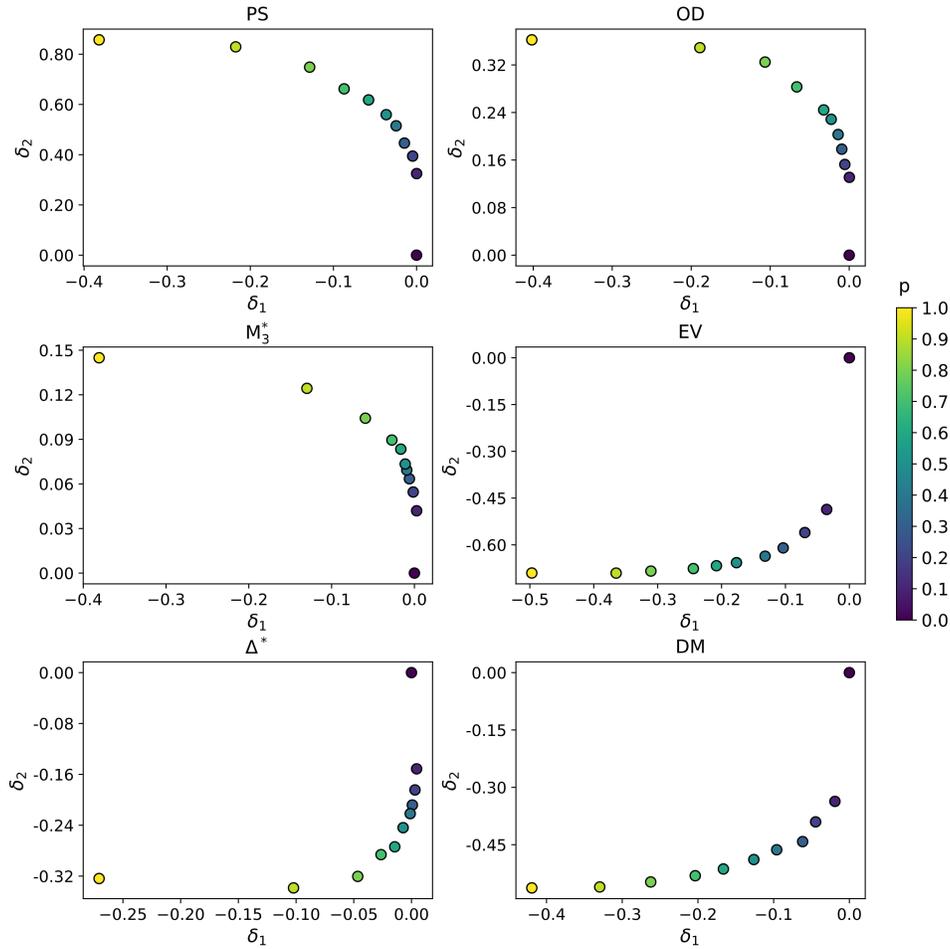}
  \centering
  \caption{
    The relative change in hypervolume ($x$-axis) and coverage ($y$-axis),
    with increasing $p$, relative to $p=0$.
    The data is based on the UMOCO-2 problem with $10$ variables and the 
    COBYLA solver. 
    Each point is the average of $20 \cdot 40$ runs 
    ($20$ instances with $40$ seeds each).
  }
  \label{fig:qmoo_vs_qmooc}
\end{figure}

As mentioned in \Cref{sec:spread}, $\QMOOC$ uses as its cost function 
a weighted average,
\begin{align*}
  p I(\Y(\theta)) + (1-p)(-\HV(\Y(\theta))),
\end{align*}
where $\Y(\theta) \subseteq \R^K$ is a Pareto front approximation generated by the ansatz
$U(\theta)$,
$p \in [0,1]$ and $\I$ is a coverage indicator.
In our experiment, we varied $\I$ in the set
\begin{align*}
  I_C := \menge{\PS, \OD, \M, \DM, \D, \EV}
\end{align*}
and $p$ in the set 
\begin{align*}
  P_C := \menge{\tfrac{1}{10}, \tfrac{2}{10}, \ldots, \tfrac{9}{10}, 1},
\end{align*}
then evaluated how coverage and hypervolume change compared to QMOO.
To be more precise, for $(I,p) \in I_C \times P_C$, let
$\Y_{I,p}$ be an output of $\QMOOC$ and $\Y$ an output of QMOO.
Since the above-mentioned indicators can vary in magnitude,
we were interested in the change of $\HV(\Y_{I,p})$, resp. $\I(\Y_{I,p})$, relative
to $\HV(\Y)$, resp. $\I(\Y)$,
\begin{align*}
  \delta_1(\HV,p) := \frac{\HV(\Y_{I,p})-\HV(\Y)}{\HV(\Y)}, \quad
  \delta_2(I,p) := \frac{\I(\Y_{I,p}) - \I(\Y)}{\I(\Y)}.
\end{align*}
\Cref{tab:qmoo_qmooc} gives an overview of our experimental setup.
\Cref{fig:qmoo_vs_qmooc} shows the average value (over multiple runs) of 
$(\delta_1(I,p), \delta_2(I,p))$ for different combinations $(I,p) \in \I_C \times P_C$.
Each subfigure corresponds to a particular $I \in \I_C$ and consists of $11$ points,
one for each hyperparameter combination in 
\begin{align*}
  \Menge{ (I,p)}{p \in P_C} \cup \menge{(I,0)}.
\end{align*}
Each of these $11$ points is the average of 
$(\delta_1(I,p), \delta_2(I,p))$ over $800 = 20 \cdot 40$ runs of QMOOC 
($20$ problem instances with $40$ seeds each).
  
\begin{table}[htbp]    
  \small       
  \centering                  
  \begin{tabular}{l|cc}
    \toprule              
     & QMOO & QMOOC \\
    \midrule
    Ansatz & $U(\theta)$ & $U(\theta)$ \\
    Number of layers $L$ & $5$ & $5$ \\
    Initial parameters & sampled via $U(0,2\pi)$ & sampled via $U(0,2\pi)$ \\
    Number of solutions $P$ & 10 & 10 \\
    Function to minimize & $-\HV(\Y(\theta))$ & 
    $pI(\Y(\theta))+(1-p)(-\HV(\Y(\theta)))$ \\
    Coverage indicator $\I$ & not relevant & $\PS, \OD, \M, \DM, \D, \EV$ \\
    Weight $p$ & not relevant & $0.1, 0.2, \ldots,1$ \\
    Classical opzimizer & \shortstack{COBYLA, Nelder-Mead, \\Powell} & 
    \shortstack{COBYLA, Nelder-Mead, \\Powell} \\
    \bottomrule
  \end{tabular}
  \caption{Overview of the setup used in comparing QMOO and QMOOC.}
  \label{tab:qmoo_qmooc}
\end{table}

We discuss the results based on the UMOCO-2 problem using the COBYLA solver. 
The other problem types and  solvers behave similarly and can be found in the appendix.

The first observation is that a coverage 
indicator may exhibit either a positive or negative 
correlation with the hypervolume.

Beyond the algorithmic influence, this relationship generally also depends on the
geometric characteristics of the Pareto front. 
Therefore, any prior knowledge about the structure of the Pareto front should
be considered when selecting a suitable indicator.
Our results indicate that, for metrics such as Pareto spread, outer diameter, 
and $\M$, 
a tradeoff between hypervolume and the coverage indicator is, 
in principle, possible.

It is interesting to note that the "tradeoff curve" for these indicators is concave, 
with a sharp rise near zero and a flatter slope the further away one moves 
to the left.
This means that one can increase coverage significantly with only 
minor sacrifices in hypervolume, after which diminishing returns set in.
For example, with Pareto Spread, coverage can be increased by $40\%$ 
without any significant loss in hypervolume.
With Outer Diameter, the increase is up to 20\%,
and with $\M$, up to 10\% at no or only minimal loss of hypervolume.

With all three indicators mentioned above, it can be observed that after 
sacrificing approximately 10\% of the hypervolume, 
the returns in the coverage indicator level off and a tradeoff appears 
less lucrative.

\begin{figure}[ht]
  \includegraphics[scale=0.096]{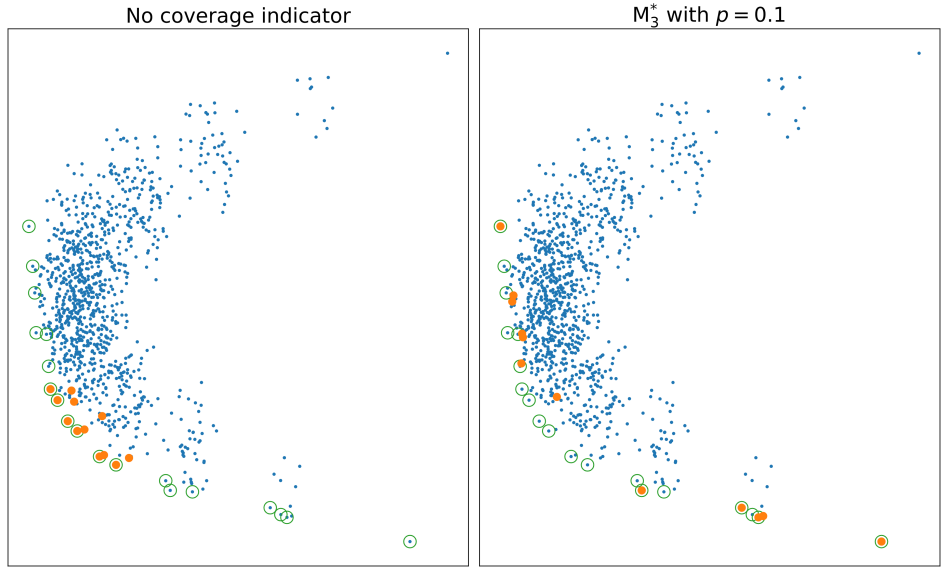}
  \centering
  \caption{
    The influence of coverage indicator $\M$ on the coverage of the Pareto front
    approximation. 
    The blue dots are the elements of $f(\F[n])$, where $f$ is an 
    instance of the problem type AFM on $10$ variables. 
    The dots circled in green form the Pareto front.
    The orange dots are the output of QMOO on the left and the output of 
    $\mathrm{QMOOC}(\M,0.1)$ on the right.
  }
  \label{fig:spread_comparison}
\end{figure}

\Cref{fig:spread_comparison} illustrates, based on an AFM example, 
how even minimal consideration of a coverage indicator 
can heaviliy influence the coverage of a Pareto front approximation.

The experiment shows that the variational quantum algorithm for 
multicriteria problems can indeed take into account coverage properties
of the Pareto front approximation without sacrificing much approximation quality. 
This is of great interest in concrete applications where an overview of different 
tradeoff options is important.

\section{Conclusion and Further Research}\label{sec:conclusion}

This work has shown that classical multicriteria algorithms can be woven 
into the variational-quantum loop to yield stronger multicriteria performance
than state of the art single-objective solver. 
By interpreting the states generated by a VQA ansatz as a multicriteria objective, 
we reformulated the parameter-search itself as a multicriteria optimization problem. 
The resulting QMOOM framework leverages population-based algorithms such as NSGA-II 
to steer the quantum circuit and, across 4 benchmark families, two qubit scales,
and 1600
statistically independent runs, delivered up to 10 percentage-point gains in
mean hyper-volume while simultaneously tightening worst-case performance.

Building on this foundation, we introduced coverage-related cost functions
that augment the traditional hypervolume objective with coverage-oriented indicators. 
A simple weighted blend was sufficient to raise Pareto front coverage by up to 40\%
with $\leq 2\%$ sacrifice in hyper-volume, 
demonstrating that diversity and quality need not be antagonistic in quantum 
optimization.

The implications are, that the multiobjective-based optimization of circuit 
parameters opens a new "plug-and-play-interface",
every classical algorithm from any metaheuristic (for example evolutionary or 
swarm methods) can in principle replace NSGA-II without modifying the circuit.

For the practitioners point of view, 
better-covered Pareto fronts translate into more informed tradeoff 
decisions in many areas of industry (for example logistics, finance)
without lengthening circuits or increasing shot counts.

Naturally, there are some limitations. 
All experiments relied on noiseless simulations with 
$n \leq 1$3 qubits and 
$K=2$ objectives; 
hyper-parameters such as layer depth, shot count and size of the
Pareto front were fixed a priori, and real-hardware validation remains pending.

Looking forward, three research topics emerge as immediate priorities. 
First, running QMOO-M and QMOOC on noisy quantum hardware, augmented 
with error-mitigations, will reveal how the coverage gains 
survive realistic noise and limited coherence. 

Second, broadening the framework to higher-dimensional objective spaces
and constrained combinatorial tasks 
(e.g. quadratic-knapsack or scheduling with hard deadlines) 
will test the generality of the meta-heuristic approach. 

Third, a theoretical analysis of convergence and sample complexity under 
common noise models could provide provable bounds, 
guiding parameter-setting and clarifying when hybrid population search
truly beats scalarised cost-functions.

By marrying quantum-state evolution with classical population dynamics, 
we take a first step toward practical, coverage-related quantum optimization,
and lay a clear roadmap for scaling the idea from simulated prototypes 
to fault-tolerant machines.

\bibliographystyle{amsalpha}
\bibliography{references}

\newpage
\appendix
\section{Appendix}


 

\begin{figure}[ht]
  \includegraphics[scale=0.33]{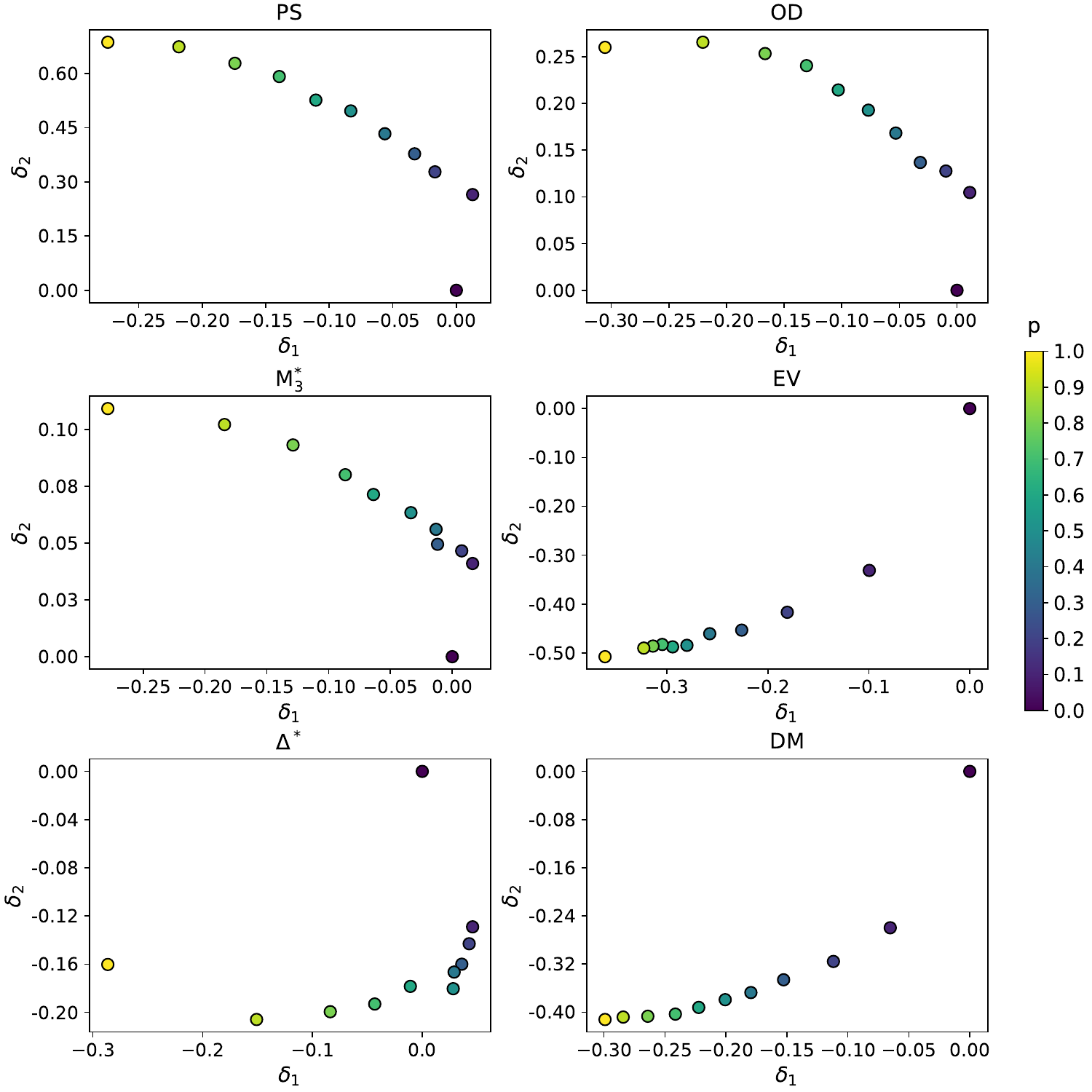} 
  \centering
  \caption{$(n, \text{problem}, \text{solver}) = (10, \text{UMOCO-2}, \text{Nelder-Mead})$.}
\end{figure}

\begin{figure}[ht]
  \includegraphics[scale=0.33]{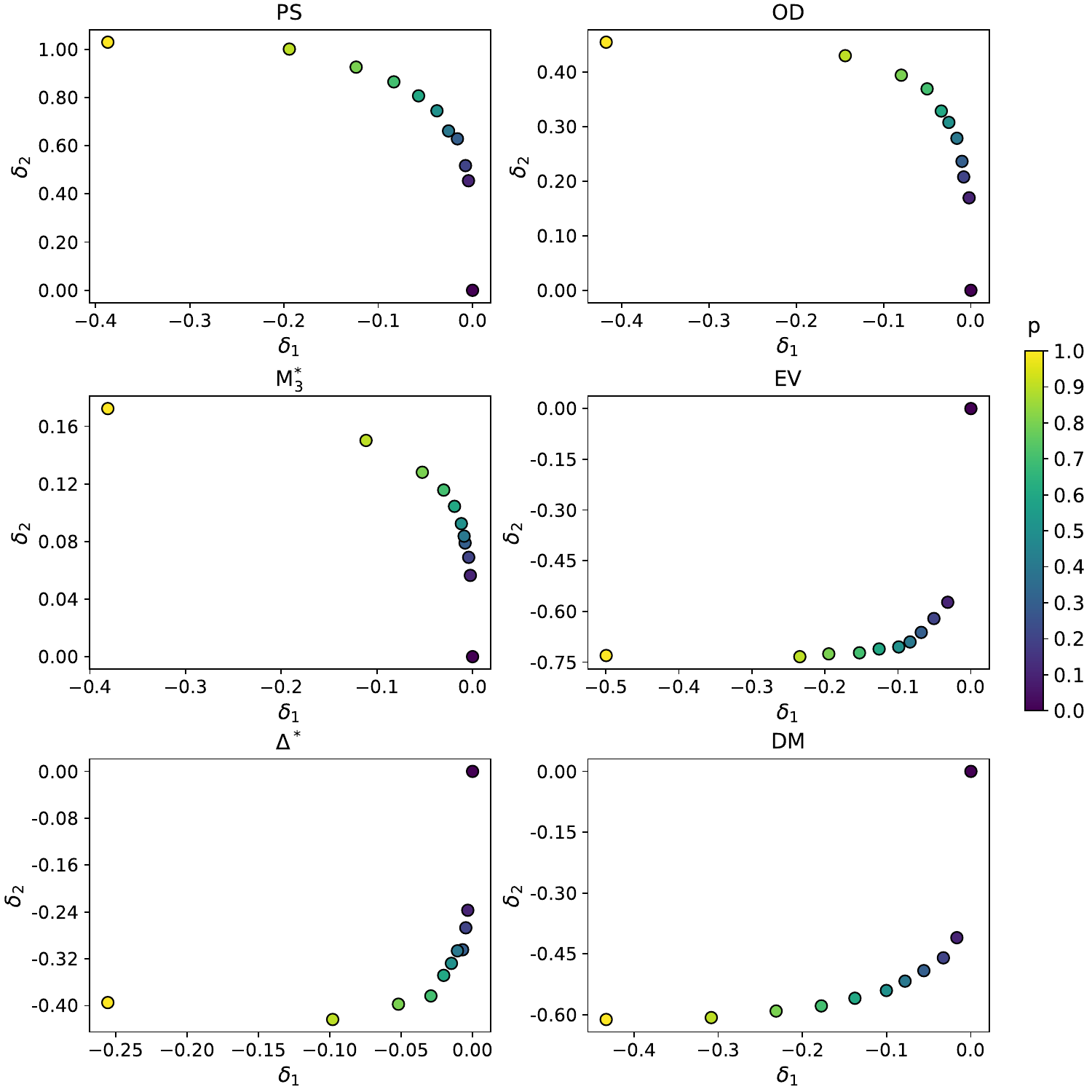} 
  \centering
  \caption{$(n, \text{problem}, \text{solver}) = (10, \text{UMOCO-2}, \text{Powell})$.}
\end{figure}

\begin{figure}[ht]
  \includegraphics[scale=0.33]{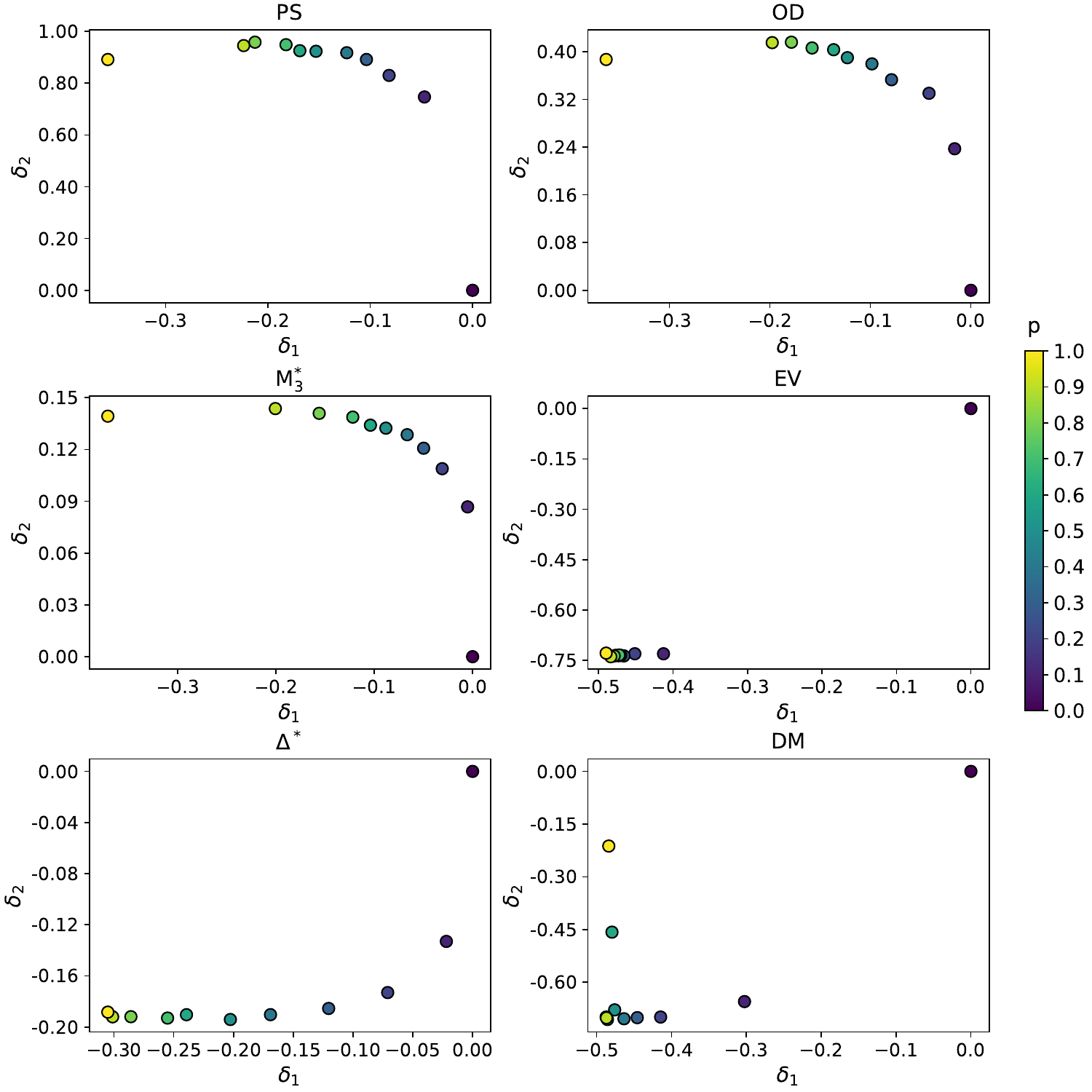} 
  \centering
  \caption{$(n, \text{problem}, \text{solver}) = (10, \text{FM-AFM}, \text{COBYLA})$.}
\end{figure}

\begin{figure}[ht]
  \includegraphics[scale=0.33]{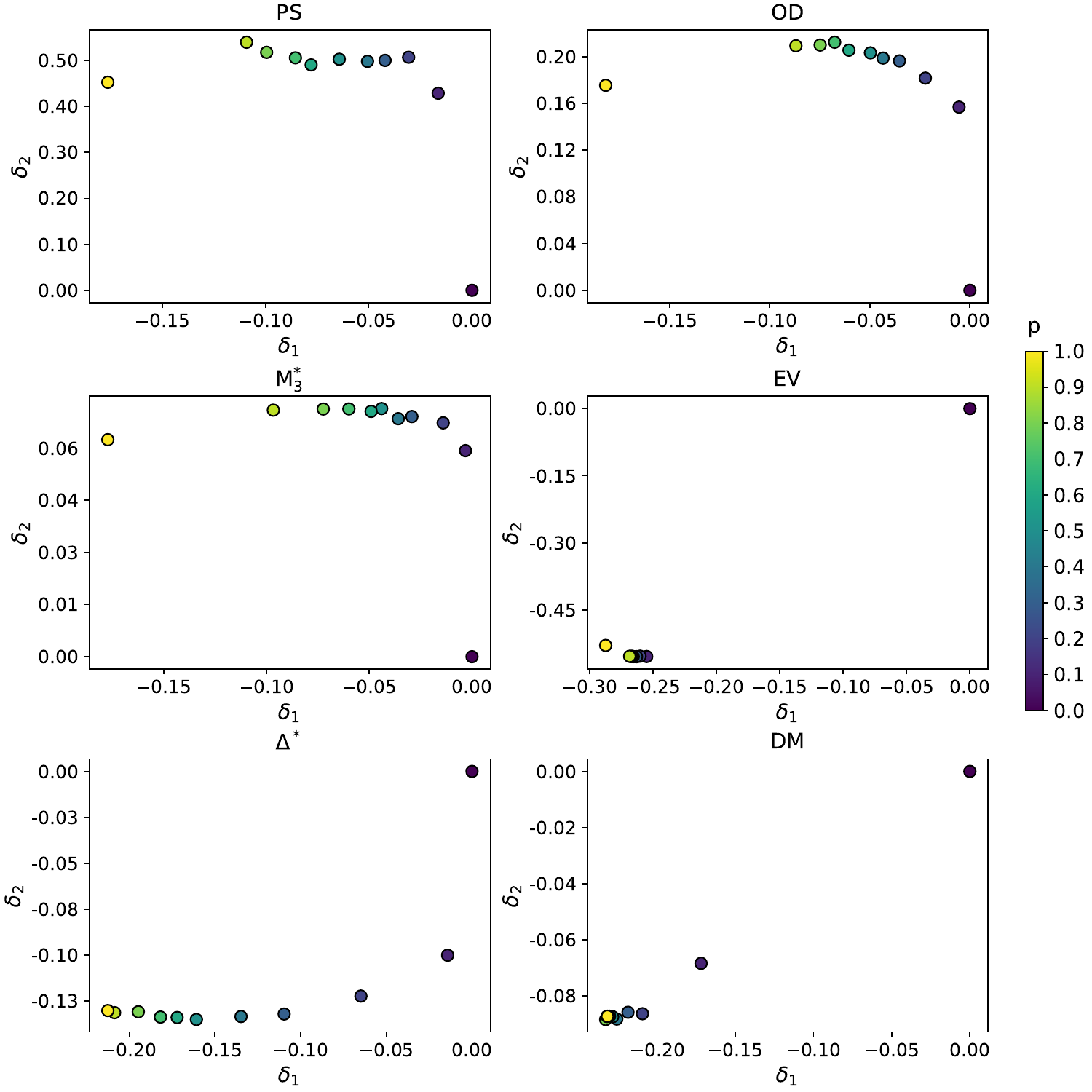} 
  \centering
  \caption{$(n, \text{problem}, \text{solver}) = (10, \text{FM-AFM}, \text{Nelder-Mead})$.}
\end{figure}

\begin{figure}[ht]
  \includegraphics[scale=0.33]{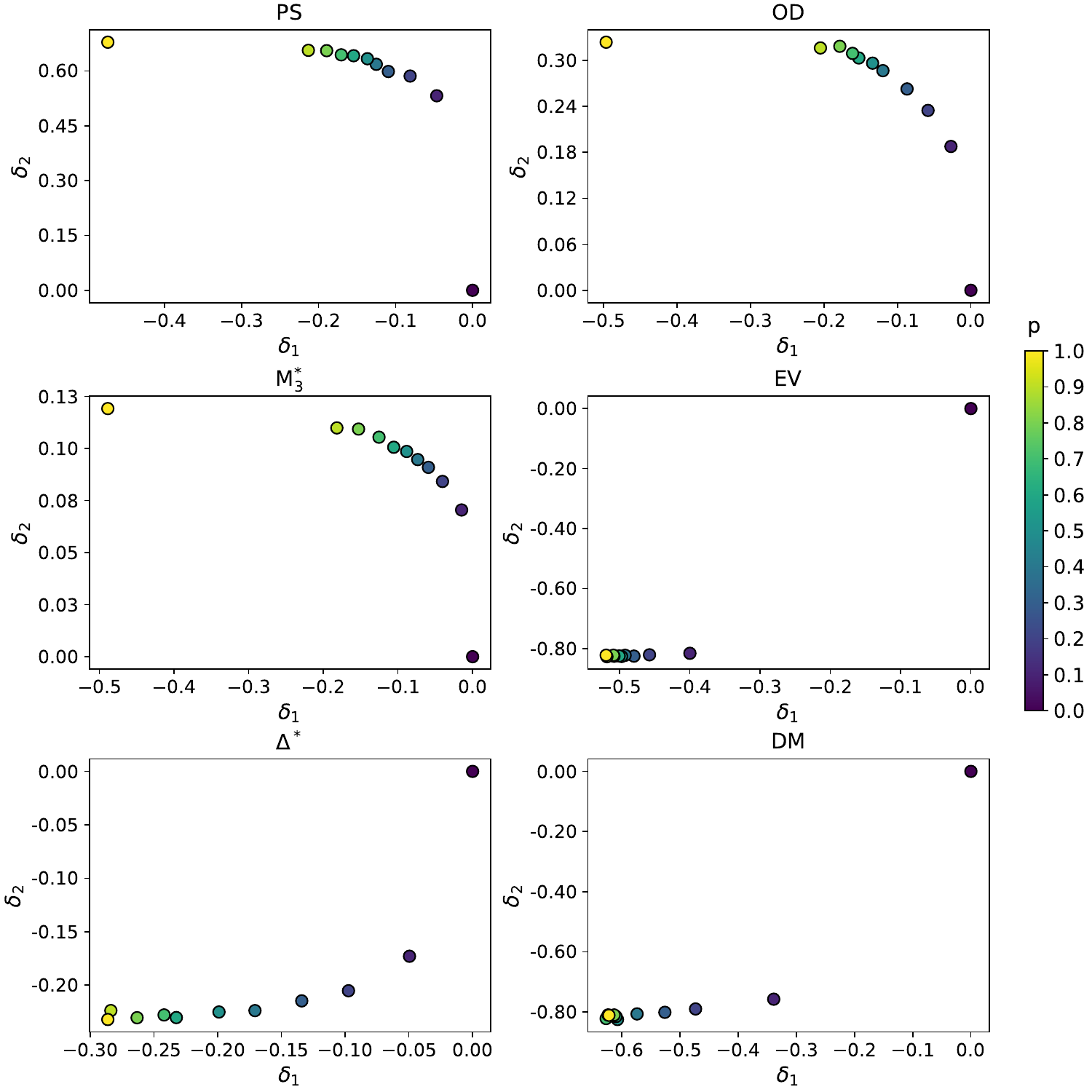} 
  \centering
  \caption{$(n, \text{problem}, \text{solver}) = (10, \text{FM-AFM}, \text{Powell})$.}
\end{figure}

\begin{figure}[ht]
  \includegraphics[scale=0.33]{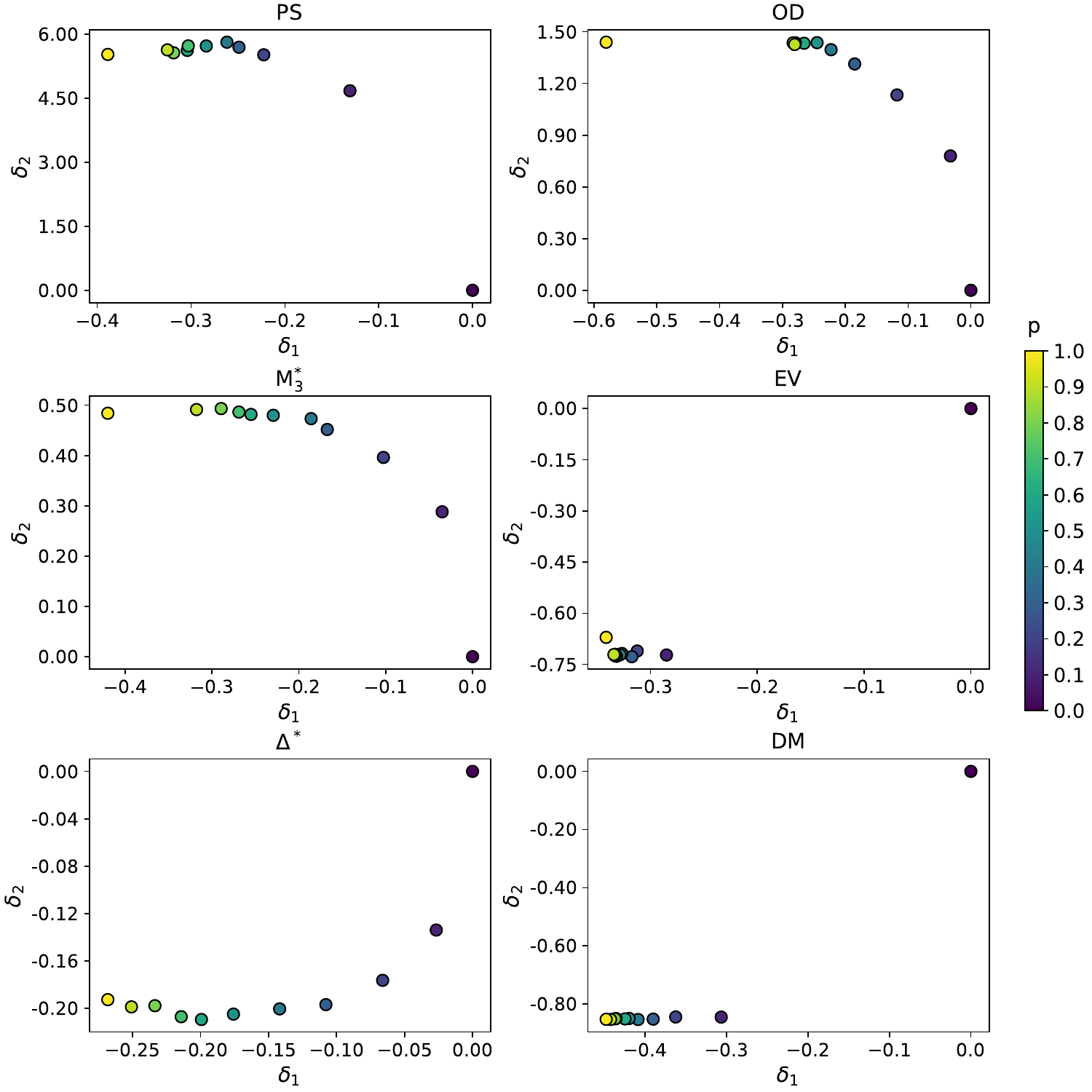} 
  \centering
  \caption{$(n, \text{problem}, \text{solver}) = (10, \text{AFM}, \text{COBYLA})$.}
\end{figure}

\begin{figure}[ht]
  \includegraphics[scale=0.33]{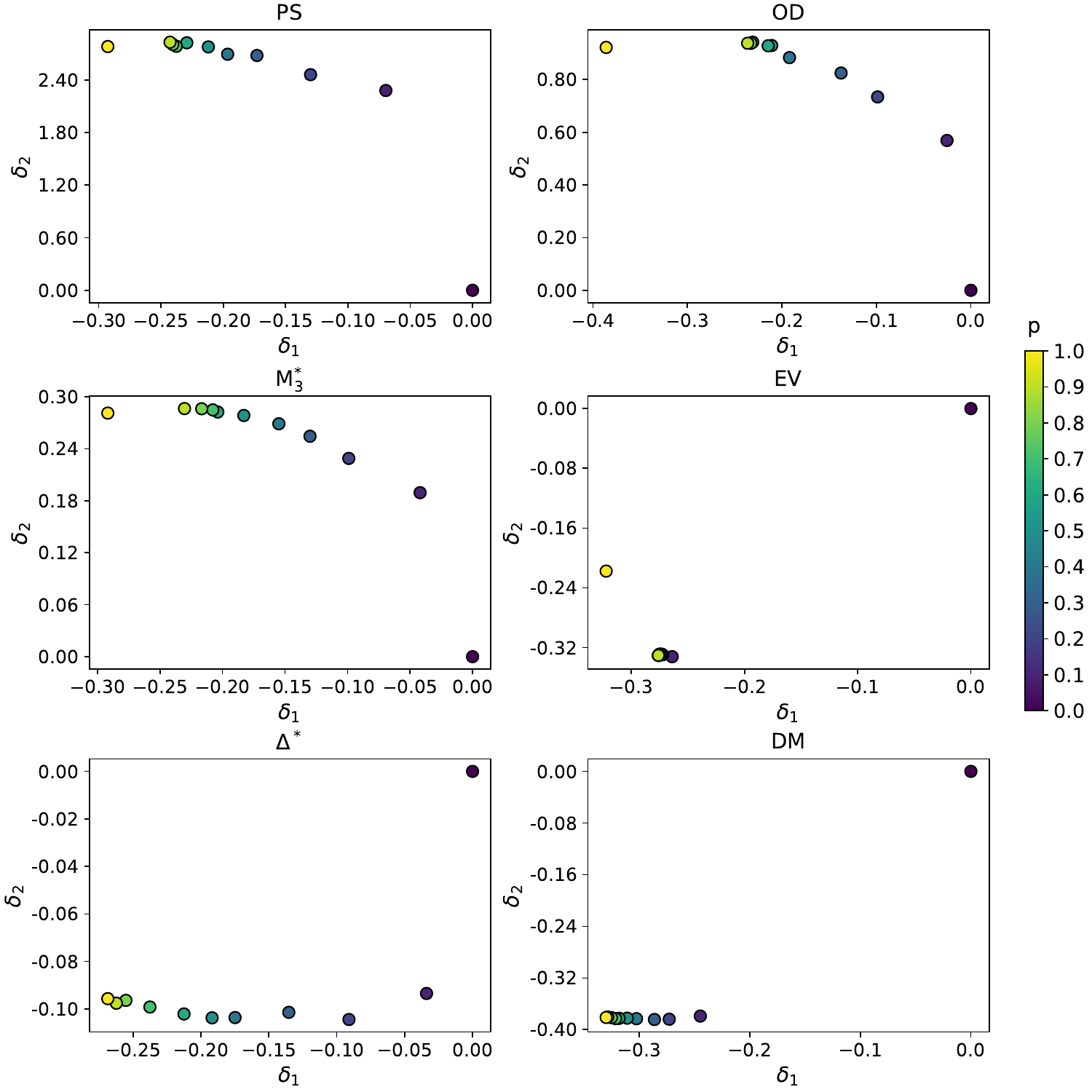} 
  \centering
  \caption{$(n, \text{problem}, \text{solver}) = (10, \text{AFM}, \text{Nelder-Mead})$.}
\end{figure}

\begin{figure}[ht]
  \includegraphics[scale=0.33]{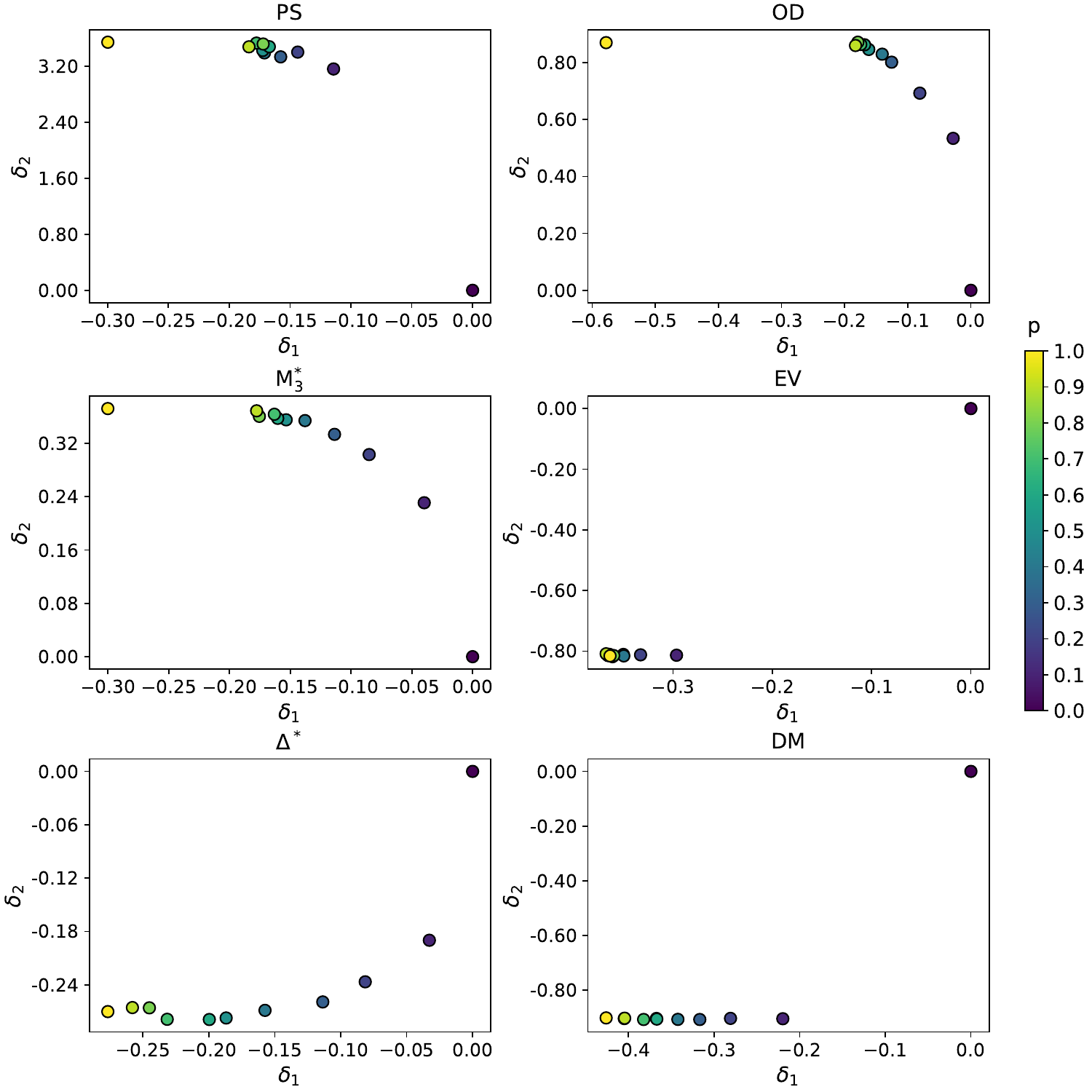} 
  \centering
  \caption{$(n, \text{problem}, \text{solver}) = (10, \text{AFM}, \text{Powell})$.}
\end{figure}

\begin{figure}[ht]
  \includegraphics[scale=0.33]{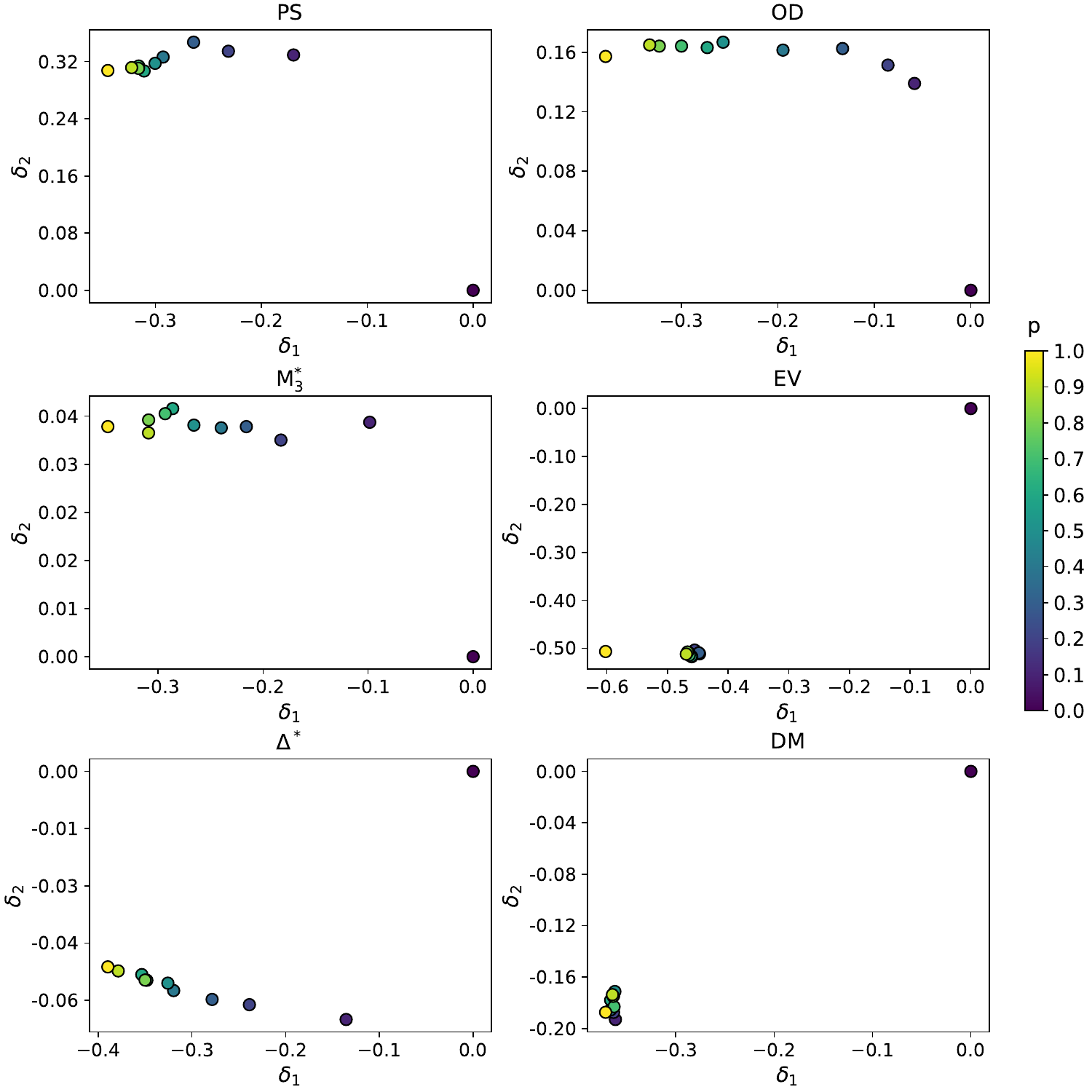} 
  \centering
  \caption{$(n, \text{problem}, \text{solver}) = (10, \text{UMOCO-1}, \text{COBYLA})$.}
\end{figure}

\begin{figure}[ht]
  \includegraphics[scale=0.33]{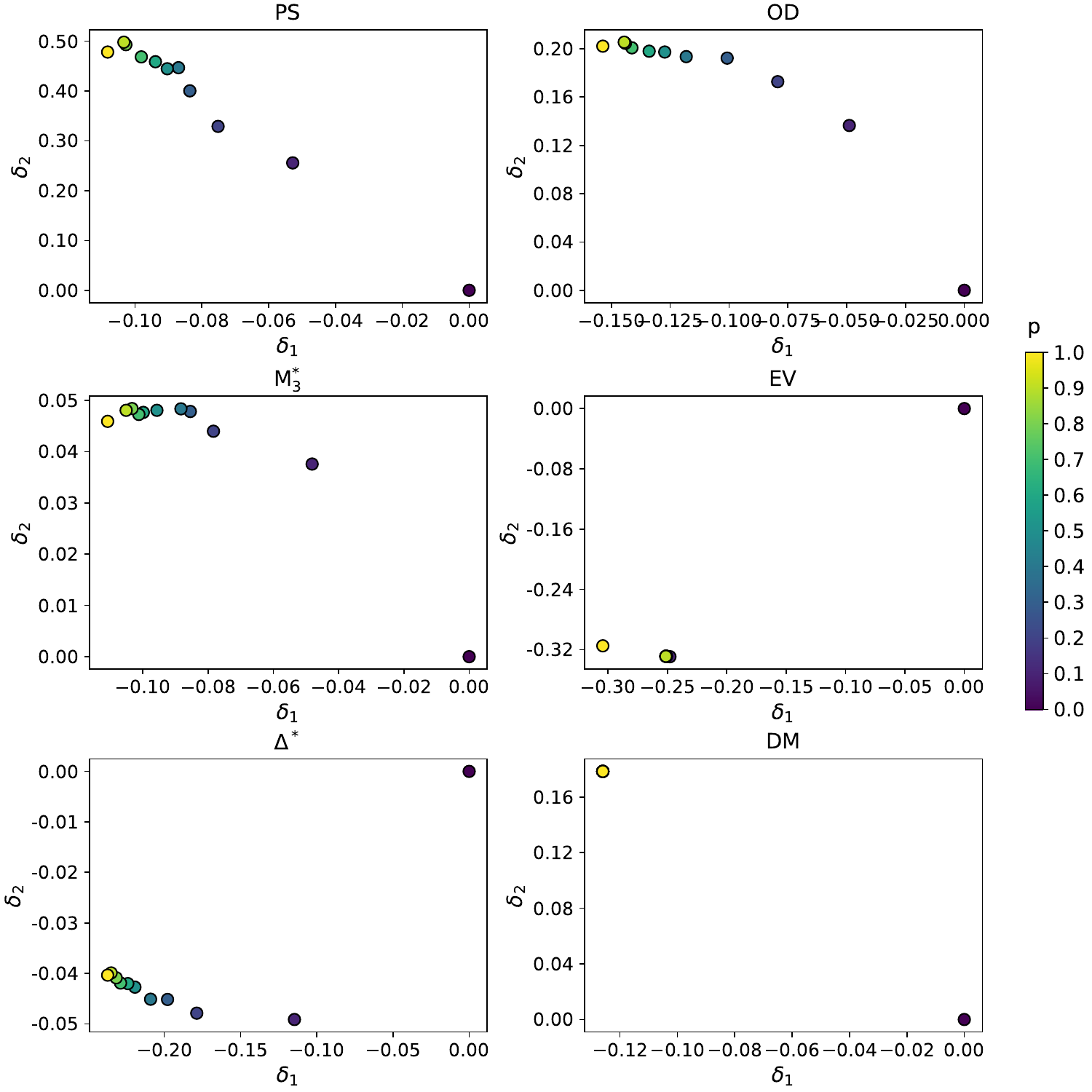} 
  \centering
  \caption{$(n, \text{problem}, \text{solver}) = (10, \text{UMOCO-1}, \text{Nelder-Mead})$.}
\end{figure}

\begin{figure}[ht]
  \includegraphics[scale=0.33]{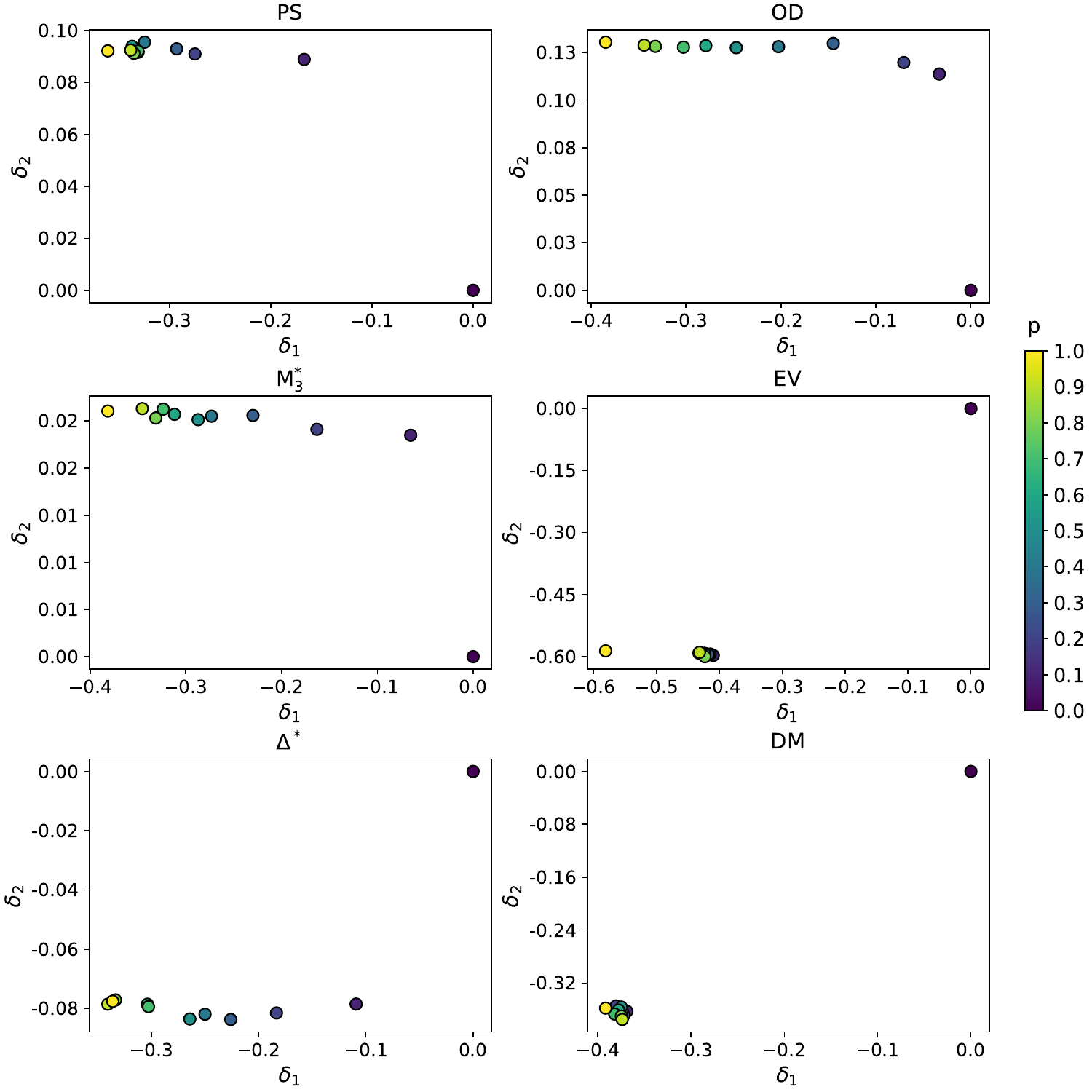} 
  \centering
  \caption{$(n, \text{problem}, \text{solver}) = (10, \text{UMOCO-1}, \text{Powell})$.}
\end{figure}

\begin{figure}[ht]
  \includegraphics[scale=0.33]{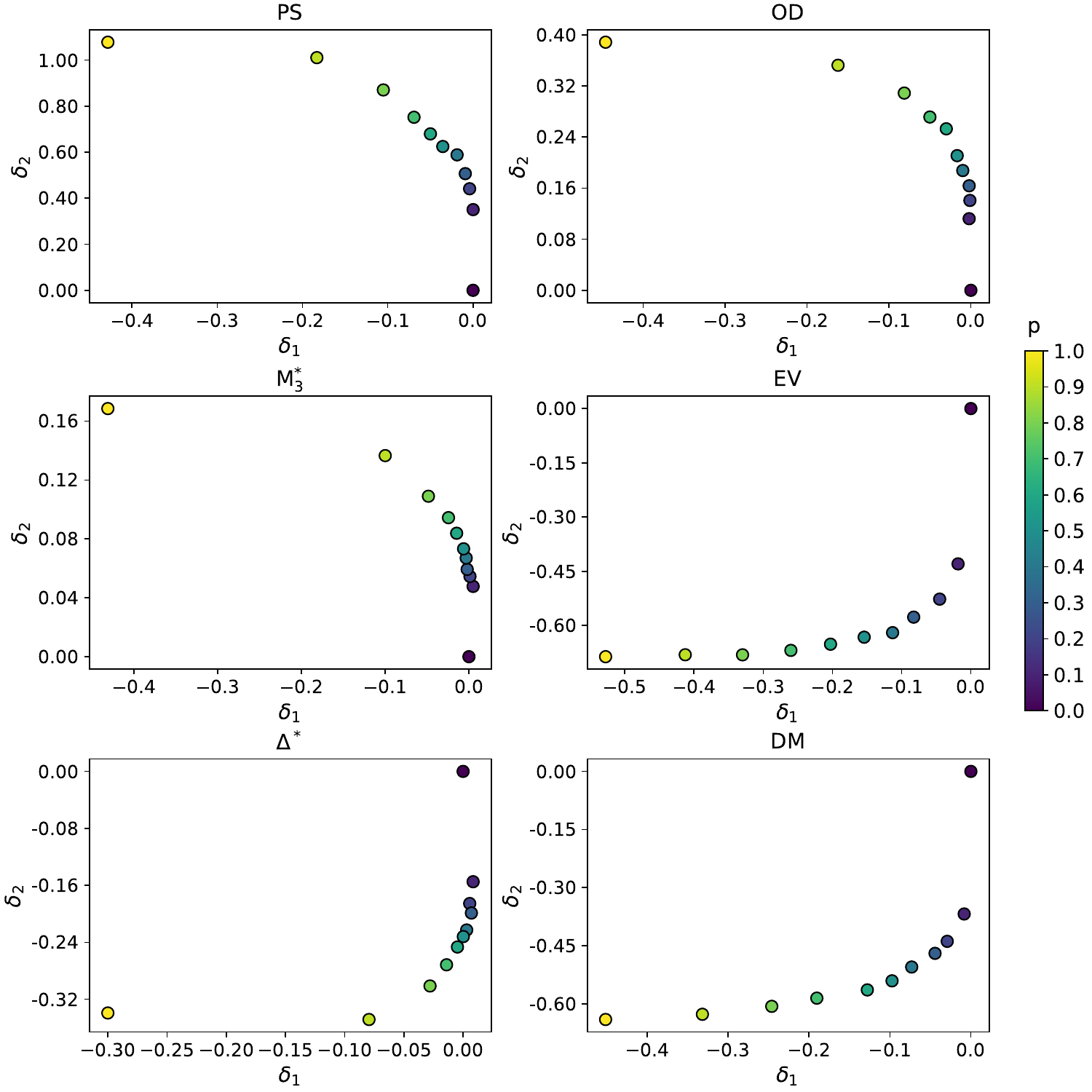} 
  \centering
  \caption{$(n, \text{problem}, \text{solver}) = (13, \text{UMOCO-2}, \text{COBYLA})$.}
\end{figure}

\begin{figure}[ht]
  \includegraphics[scale=0.33]{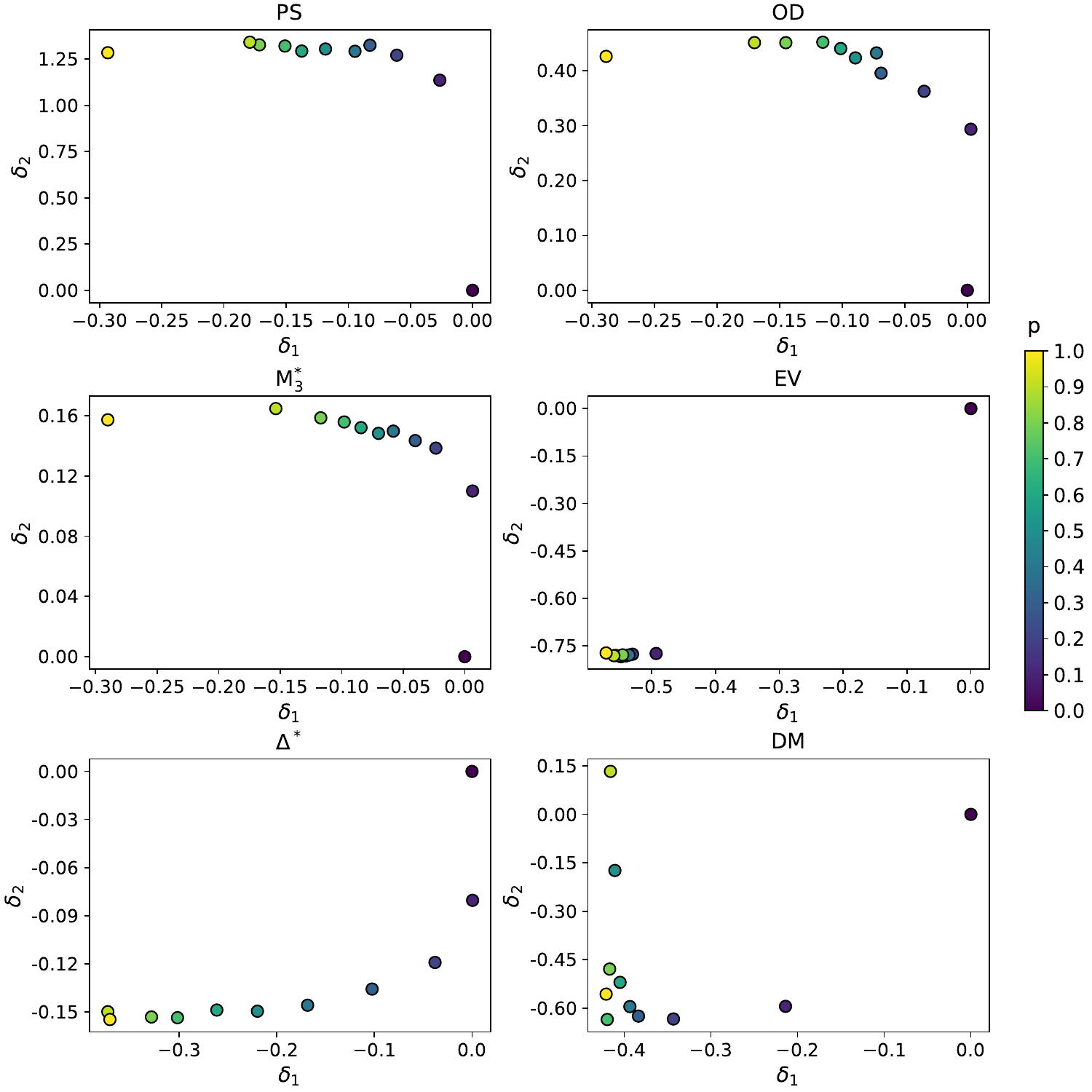} 
  \centering
  \caption{$(n, \text{problem}, \text{solver}) = (13, \text{FM-AFM}, \text{COBYLA})$.}
\end{figure}

\begin{figure}[ht]
  \includegraphics[scale=0.33]{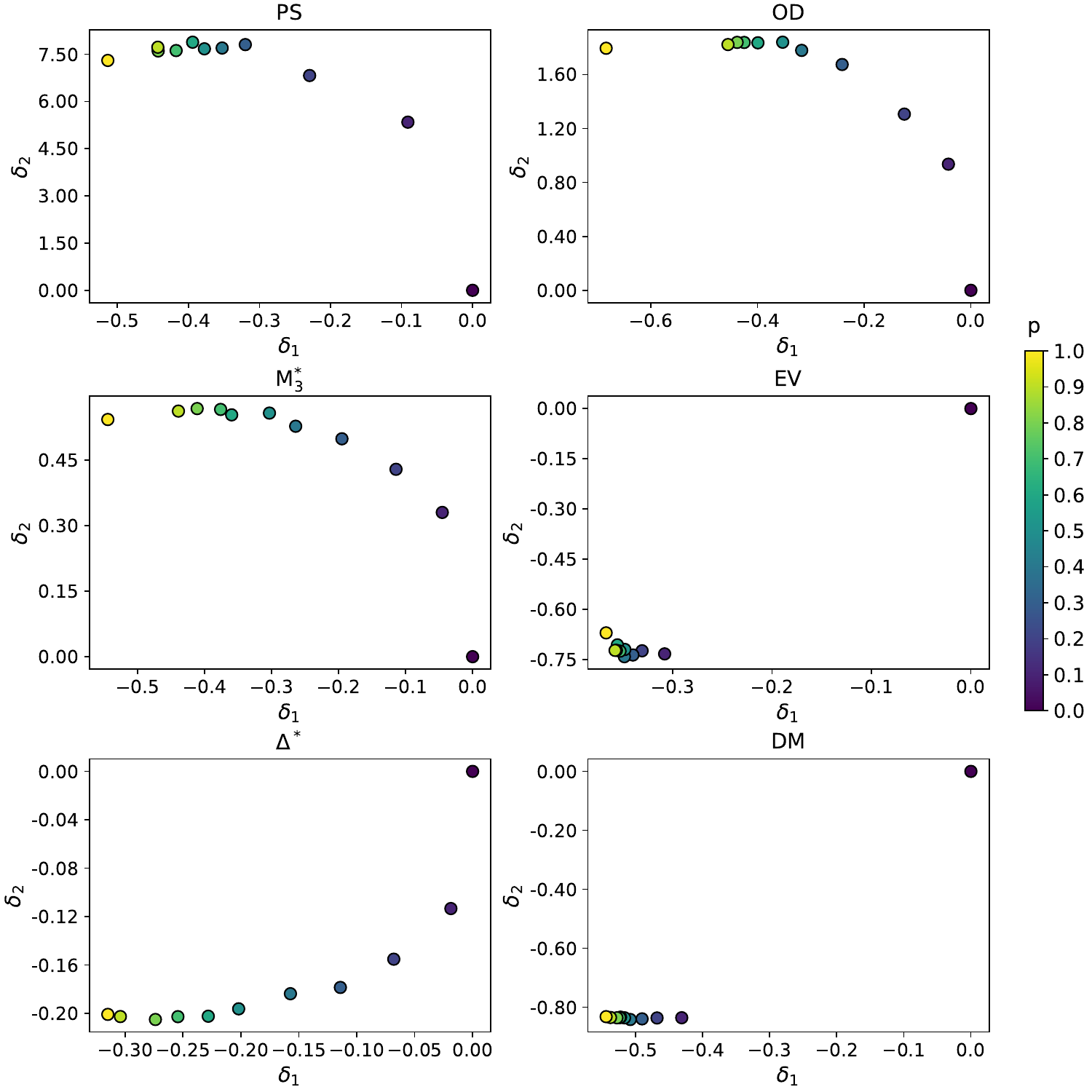} 
  \centering
  \caption{$(n, \text{problem}, \text{solver}) = (13, \text{AFM}, \text{COBYLA})$.}
\end{figure}

\begin{figure}[ht]
  \includegraphics[scale=0.33]{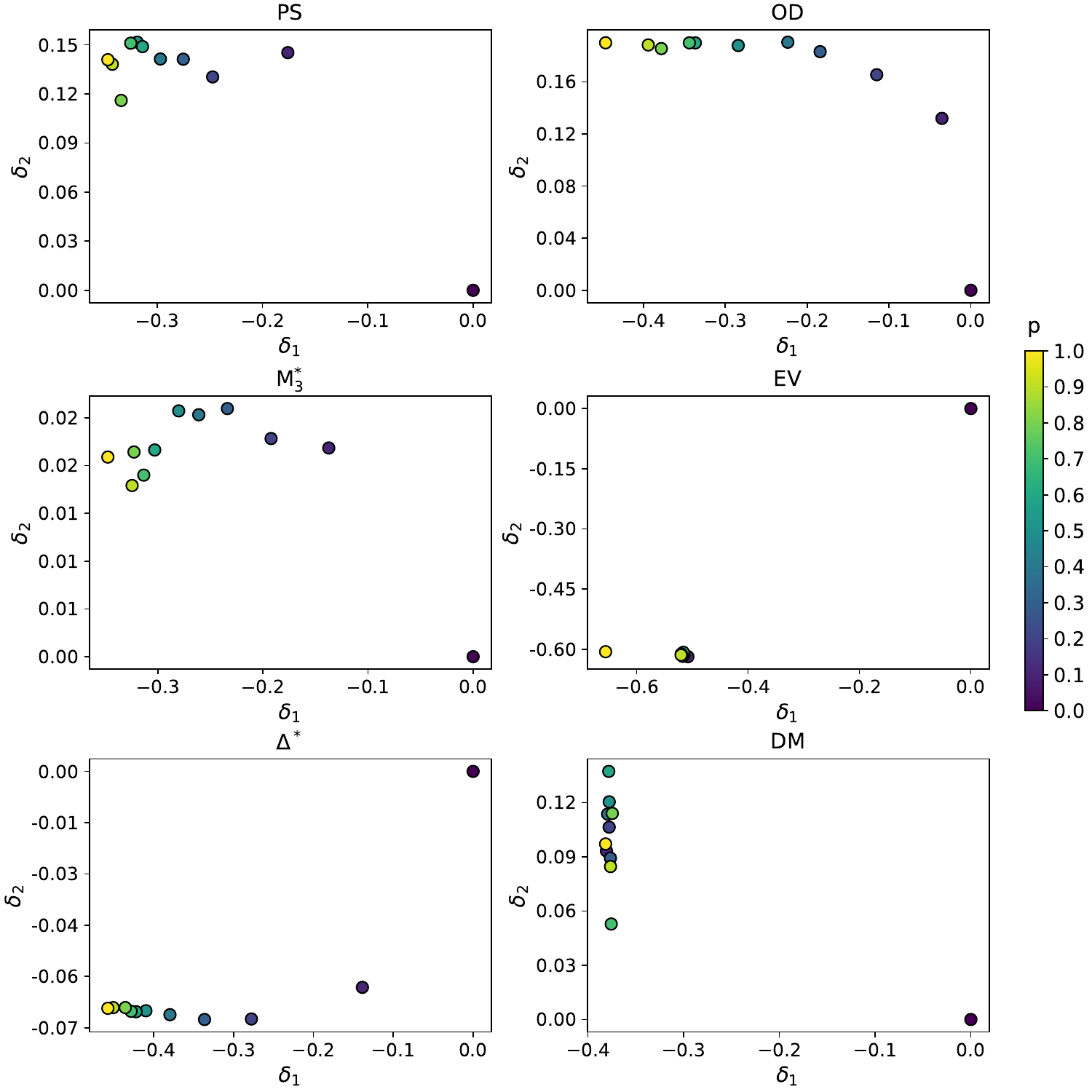} 
  \centering
  \caption{$(n, \text{problem}, \text{solver}) = (13, \text{UMOCO-1}, \text{COBYLA})$.}
\end{figure}

\end{document}